\documentclass[journal]{IEEEtran}

\usepackage{booktabs} 
\usepackage[T1]{fontenc}
\usepackage[utf8]{inputenc}

\usepackage{tikz}
\usetikzlibrary{shapes,arrows.meta,shadows}
\usepackage{graphicx}

\usepackage{amsmath,bm}
\usepackage{paralist}
\usepackage{amsfonts}
\usepackage{amssymb}
\usepackage{array}
\usepackage{amsthm}
\newtheorem{theorem}{Theorem}
\newtheorem{proposition}[theorem]{Proposition}
\newtheorem{lemma}[theorem]{Lemma}
\newtheorem{corollary}[theorem]{Corollary}
\theoremstyle{definition}
\newtheorem{definition}{Definition}
\theoremstyle{remark}
\newtheorem*{remark}{Remark}
\newcommand{\E}{\mathbb{E}}

\newcommand{\Prb}{\mathbb{P}}

\newcommand{\Indc}{\mathcal{I}}
\newcommand{\cntrl}{p}
\newcommand{\gacq}{\gamma}

\usepackage{comment}
\usepackage{url}
\usepackage{subcaption}
\usepackage{multirow}

\usepackage{accents}
\usepackage{environ}
\makeatletter
\newsavebox{\measure@tikzpicture}
\NewEnviron{scaletikzpicturetowidth}[1]{%
  \def\tikz@width{#1}%
  \def\tikzscale{1}\begin{lrbox}{\measure@tikzpicture}%
  \BODY
  \end{lrbox}%
  \pgfmathparse{#1/\wd\measure@tikzpicture}%
  \edef\tikzscale{\pgfmathresult}%
  \BODY
}
\makeatother
\tikzset{>={Latex[width=3mm,length=3mm]}}
\DeclareMathOperator*{\argmax}{argmax}
\DeclareMathOperator*{\argmin}{argmin}

\usepackage[font={small}]{caption}

\newsavebox{\smlmat}
\savebox{\smlmat}{$\begin{bmatrix}0.8&0.2\\0.4&0.6\end{bmatrix}$}
\def\BibTeX{{\rm B\kern-.05em{\sc i\kern-.025em b}\kern-.08em
		T\kern-.1667em\lower.7ex\hbox{E}\kern-.125emX}}

\title{Controlled Sequential Information Fusion with Social Sensors}

\author{\hspace{4cm}~Sujay~Bhatt~~~
        Vikram~Krishnamurthy,~\IEEEmembership{Fellow,~IEEE}
\thanks{S. Bhatt and V. Krishnamurthy are with the Department
of Electrical and Computer Engineering, Cornell University, Ithaca,
NY, 14853 USA.}  
\thanks{E-mail: (sh2376@cornell.edu), (vikramk@cornell.edu).}
\thanks{Partial results appear in ACM SIGMETRICS Performance Evaluation Review, 2017, New York, NY, USA.}}

\begin{document}
\maketitle

\begin{abstract}
A sequence of social sensors estimate an unknown parameter (modeled as a state of nature) by performing Bayesian Social Learning, and myopically optimize individual reward functions. The decisions of the social sensors contain quantized information about the underlying state. How should a fusion center dynamically incentivize the social sensors for acquiring information about the underlying state? 

This paper presents five results. First, sufficient conditions on the model parameters are provided under which the optimal policy for the fusion center has a threshold structure. The optimal policy is determined in closed form, and is such that it switches between two exactly specified incentive policies at the threshold. Second, it is shown that the optimal incentive sequence is a \textit{sub-martingale}, i.e, the optimal incentives increase on average over time. Third, it is shown that it is possible for the fusion center to learn the true state asymptotically by employing a sub-optimal policy; in other words, controlled information fusion with social sensors can be consistent. Fourth, uniform bounds on the average additional cost incurred by the fusion center for employing a sub-optimal policy are provided. This characterizes the trade-off between the cost of information acquisition and consistency for the fusion center. Finally, when it is sufficient to estimate the state with a degree of confidence, uniform bounds on the budget saved by employing policies that guarantee state estimation in finite time are provided.
\end{abstract}


%
\IEEEpeerreviewmaketitle

\section{Introduction}

A social sensor\footnote{A social (human) sensor provides information about its state (sentiment, social situation, quality of product) to a social network after interaction with other social sensors. In this paper, in line with a large body of literature, we adopt a more stylized definition: a \textit{social sensor performs social learning}.} is an information processing system that differs from a physical sensor in the following ways:
\begin{compactenum}
\item[i.)] Social sensors influence the behavior of other sensors, whereas physical sensors typically do not affect other sensors.
\item[ii.)] Social sensors reveal quantized information (decisions) and have dynamics, whereas physical sensors are static with the dynamics modeled in the state equation.
\end{compactenum}
Social learning is the process by which social sensors are influenced by the behaviour of other sensors in the social network. 

This paper considers a sequential decision making model of Bayesian social learning introduced in~\cite{BHW92,Wel92,Ban92}, where the social sensors learn from their predecessors. Each social sensor has a private signal on the underlying state and considers this in addition to the (bounded) information gathered by its predecessors. 
%
\begin{figure*}[t!]
    \begin{subfigure}[t]{0.5\textwidth}
        {\hspace{-0.2cm}}\includegraphics[width=6.5cm,height=2.5cm]{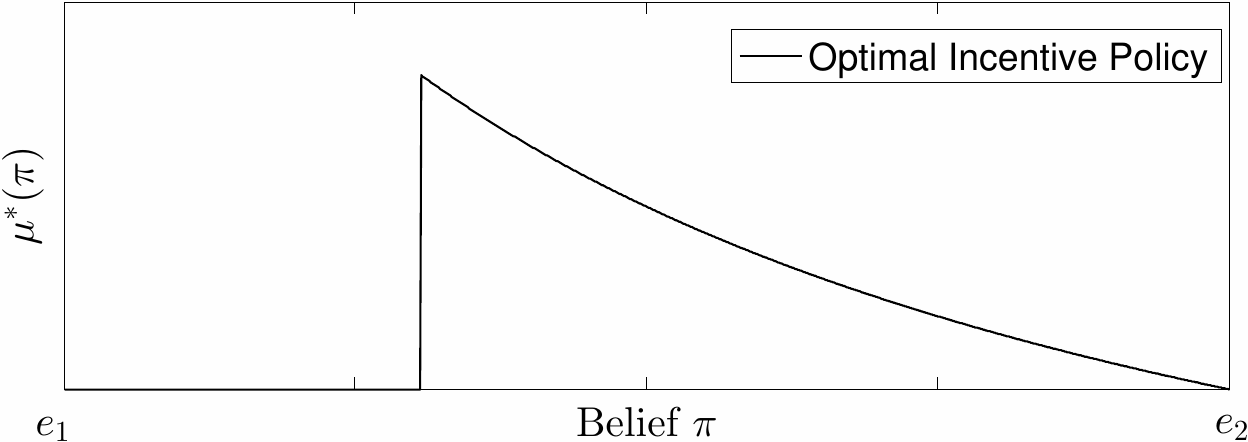}
        \caption{Single Threshold Incentive Policy}
        \label{subfig1:STMT}
    \end{subfigure}%
    ~~
    \begin{subfigure}[t]{0.5\textwidth}
       {\hspace{0.1cm}}\includegraphics[width=6.5cm,height=2.5cm]{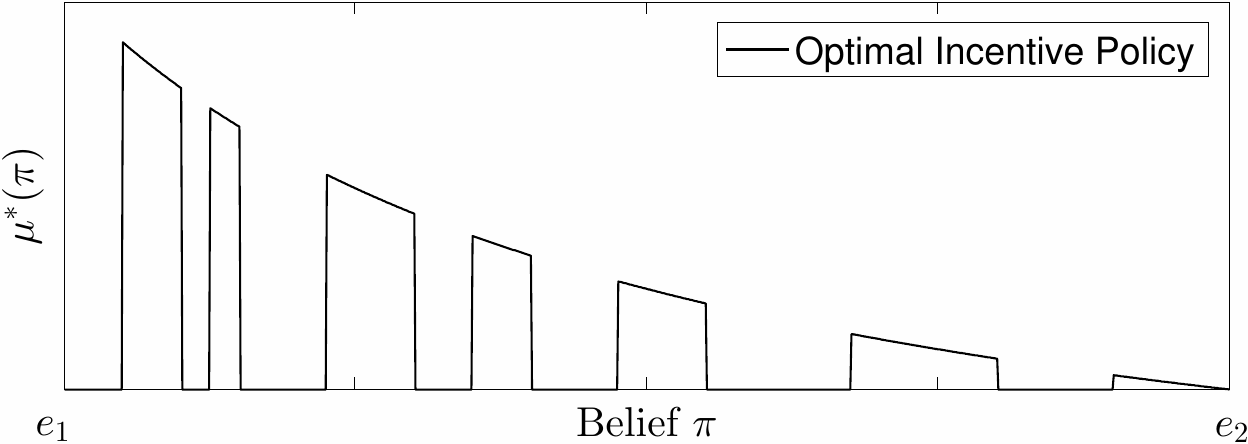}
        \caption{General (Multi-threshold) Incentive Policy}
        \label{subfig2:STMT}
    \end{subfigure}
    \caption{Visual illustration of the main result of the paper, namely the optimal incentive policy for choosing incentives to social sensors to reveal information to the fusion center. In general, the incentive policy could be an arbitrary function of $\pi$ as in Fig.1b; we provide conditions under which the incentive policy is as in Fig.1a.  Here $\mu^*(\pi) \in [0,1]$ is the optimal incentive policy and $\pi$ denotes the estimate of the state computed from the fusion of sensors' decisions. $e_1$ and $e_2$ denote the indicator vectors. When $p = \mu^*(\pi) = 0$, the fusion center should not incentivize. The optimal policy is a choice between two exactly specified incentive policies, and hence is determined in closed form. Sec.\ref{subsec:MTP} provides numerical examples (of more general cost functions) where the optimal incentive policy is multi-threshold as in Fig.\ref{subfig2:STMT}.}
    \label{fig:STMT}
\end{figure*}
\subsection*{Information Fusion with Social Sensors}
Information fusion with physical sensors is a well studied problem. In this paper, motivated by recent applications using online social media review platforms, we consider information fusion with social sensors. We consider the following problem: A sequence of social sensors estimate an unknown state of nature, and a fusion center aims to estimate the underlying state by incentivizing the social sensors. How should the fusion center dynamically incentivize the social sensors for acquiring information about the underlying state? Equivalently, how can the fusion center optimize the trade-off between the cost of information acquisition from the social sensors versus the usefulness of the information in terms of reduction in uncertainty (mean-square error between the true state and the estimate) of the Bayesian state estimate.

The problem of information fusion with social sensors considered in this paper deals with combining the decisions of the social sensors, correlated due to social learning, to make an informed decision regarding the underlying state (that is being estimated). Information fusion with social sensors is challenging due to the fact that social learning leads to inefficiencies \cite{BHW92,Ban92,Wel92} like herds (sensors choose the same action irrespective of their private information) and informational cascades (information fusion results in no improvement in uncertainty). So having more social sensors need not always be advantageous (in terms of reduced mean square error between the state estimate and the true state). 

Multi-sensor data fusion \cite{CT85,Var12,JU17,CHHY18} on the other hand, refers to the problem of data acquisition, processing, and fusion of information, to provide a better estimate of the underlying state. A data fusion center gathers the information from the peripheral sensors (physical sensors) to make an informed decision regarding the desired parameter. Having more number of sensors leads to improvement in reliability, resolution, coverage, and confidence; see~\cite{Var12}. 

Traditionally, information fusion is open-loop; in this paper, we use feedback control to choose incentives to control how the sensors provide information. Hence we name the problem considered in this paper as \textit{controlled information fusion}. The fusion is Bayesian and we are interested in designing the control laws for providing optimal incentives for social sensors that will result in accurate Bayesian estimates.


In controlled fusion, the process of incentivization modifies the cost or reward function of the social sensors and hence directly affects the sensors' decisions (see~Fig.\ref{fig:SMIP}). The decisions are a quantization of the Bayesian estimate of the state, and hence controlling the incentives can shape the information that is subsequently fused. \cite{BK18} provides partial results in this direction, namely, the single threshold policy (Theorem~\ref{thm:OPRM}) and sub-martingale property (Theorem~\ref{thm:SMP}) under more restrictive conditions on the reward function of the social sensors (see Appendix~\ref{App:Dis}). This paper extends the results to more general reward functions and also provides finite-time statistical guarantees for the derived polices (algorithms).
%
%
\subsection*{Additional Related Literature}
\cite{SS00} show that some of the inefficiencies in the Bayesian sequential social learning model -- herds and cascades -- arise due to the bounded nature of the information or beliefs used in decision making, and show that the true state is aggregated when the beliefs are unbounded. There are many works in Bayesian social learning, where the social sensors repeatedly make decisions. \cite{Orl95} considers a model of repeated decision making where each sensor considers the group opinion as a summary of all the observed decisions. It is shown that cascades need not appear and weight attached to the decision of others determines the influence of private information on decision making. \cite{BG98} consider a situation where only the payoffs as opposed to signals are observed and show that there is asymptotic learning -- beliefs converge and all social sensors herd on the same action. \cite{GK03} consider rational learning over a social network, where the social sensors not always observe the decisions of other sensors, but instead make inferences on the unseen actions. The relation between network structure and herding is established. \cite{ADLO11} establish asymptotic learning on general social network topologies and establish conditions for asymptotic learning. See also \cite{SYS17} and the references therein for social learning with repeated decision making. In case of repeated decision making, inefficiencies like cascades and herds can be avoided, however, that comes at the cost of increased computational complexity for the individual social sensors. \\ 

\subsection*{Main Results and Organization}
In the context of controlled information fusion, this paper has $3$ main topics: \\
($1.$) \textbf{Optimality of Threshold Incentive Policy}:~Sec.\ref{subsec:SOIR}, gives sufficient conditions on the model parameters under which the optimal incentive policy for the fusion center has a threshold structure (see Fig.\ref{subfig1:STMT}), when estimating a random variable. Indeed we will show that the optimal policy switches between two exactly specified incentive policies at the threshold, and hence is completely determined in closed form. Since the optimal policy is determined in closed form, the fusion center only needs to store the threshold state $\pi^*$ and the incentive function, so a threshold policy is practically useful. \\ ($2.$)~\textbf{Sub-martingale Property of Optimal Incentive Sequence}:   
While Sec.\ref{subsec:SOIR} establishes the structure of the optimal incentive policy, Sec.\ref{sec:POPS} establishes the sample path properties of the optimal incentive {\em sequence}, when estimating a random variable. In particular, we show that the optimal incentive sequence is a sub-martingale; i.e, the incentives \textit{increase} on average over time. The increase can be attributed to the fact that the senors polled for information at a later instant associate a higher value on average due to learning from their predecessors. 
This property is useful in assessing the reliability of the fusion center. In a related context, our result is similar to the super-martingale property of pricing policies in economics \cite{BOOV06,BOOV08}; which says that the optimal pricing policy for charging sensors (performing social learning) who purchase a product, is to start high, establish an elite customer base, and then decrease prices to increase profits. Sec.\ref{sec:IIFDS} illustrates the difficulty of characterizing the structure of the optimal incentive sequence when the underlying state is changing according to a Markov chain. We provide conditions on the state transitions which guarantee that the optimal policy for estimating a random variable is near optimal for tracking the Markov chain. \\ 
($3.$) \textbf{Consistency of Controlled Information Fusion}:~
Information fusion with social sensors is challenging due to the fact that social learning terminates after a finite horizon~\cite{Cha04,Kri16} due to the formation of information cascades.  We show that the inefficiencies in the sequential social learning model--herds and information cascades -- can be controlled using the incentives (Corollary~$4$) and provide uniform bounds (Theorem~$5$ and Theorem~$6$) on the performance. Previous work \cite{Cha04,ABO14, ADLO11} emphasized providing conditions on the observation distribution or action sampling distributions for the social sensors. Sec.\ref{sec:CIF} shows how the fusion center can control the incentives and learn the true state asymptotically by employing a sub-optimal policy; in other words, how to control the incentives such that the information fusion with social sensors is consistent (convergence in probability). However, by employing a sub-optimal policy, the fusion center incurs additional cost. Therefore, uniform bounds on the average additional cost incurred by the fusion center for employing a sub-optimal policy are provided. These bounds characterize the trade-off between the cost of information acquisition and consistency, for the fusion center. When it is sufficient to know the state with a degree of confidence, policies that guarantee state estimation in finite time are discussed. Uniform bounds on the budget saved as a result of estimating the state only upto a degree of confidence are provided. \\
Sec.\ref{sec:NuEx} presents numerical examples that provide additional insights on the main results. A discussion on extension to multiple actions and states is provided in Sec.\ref{subsec:CFNB}. 
\section{Social Learning Model and fusion center Objective} \label{sec:PPCO}


\pgfdeclarelayer{background}
\pgfdeclarelayer{foreground}
\pgfsetlayers{background,main,foreground}


\tikzstyle{sensor}=[draw, fill=blue!20, text width=7em, 
    text centered, minimum height=2.5em,drop shadow]
    \tikzstyle{stat}=[draw, fill=green!20, text width=7em, 
    text centered, minimum height=2.5em,rounded corners,drop shadow]
\tikzstyle{ann} = [above, text width=5em, text centered]
\tikzstyle{wa} = [sensor, text width=10em, fill=red!20, 
    minimum height=3em, rounded corners, drop shadow]
\tikzstyle{sc} = [sensor, text width=13em, fill=black!20, 
    minimum height=10em, rounded corners, drop shadow]

\def\blockdist{2.3}
\def\edgedist{2.5}
\begin{figure}
\centering
  \begin{scaletikzpicturetowidth}{0.35\textwidth}
\hspace{-0cm}  \begin{tikzpicture}[scale=\tikzscale]

    \node (wa) [sensor]  {Social Sensor~$k$};
    \path (wa.south)+(0,-1.7) node (asr1) [stat] {State~$x \sim \pi$};
   
    \path (wa.north)+(0,2) node (vote) [wa] {Information \\ Fusion Center};
    

    \path [draw, dashed, thick, ->] (asr1.north) -- node [midway,right] {Observation~$y_k$} 
        (wa.south) ;
  
     \path [draw, thick, ->]  (wa.east)-- node [above] {Decision~$a_{k}$} 
        (4.5,0) ;
  
    \path [draw, thick, ->] (vote.south) -- node [midway,right] {Incentive~$p_k = \mu(\pi_{k-1})$} 
        (wa.north);

       
       \path (wa.west) +(-1.5,1.2) node (pbr) {Public Belief~$\pi_{k-1}$}; 
       
       \path [draw, dashed, thick, ->] (pbr.north) |- node [midway,above] {} 
        (vote.west) ;
        
         \path [draw, dashed, thick, ->] (pbr.south) |- node [midway,above] {} 
        (wa.west) ;
               
    \path (wa.south) +(-0,-2.8) node (asrs) {Controlled Information Fusion};
  
    \begin{pgfonlayer}{background}
        \path (asr1.west |- asr1.north)+(-3.2,5) node (a) {};
        \path (wa.north -| wa.north)+(+4,-2) node (b) {};
        \path (vote.east |- vote.north)+(+2.4,-7) node (c) {};
          
        \path[fill=brown!20,rounded corners, draw=black!50, dashed]
            (a) rectangle (c);           
        \path (asr1.north west)+(-3,3) node (a) {};
            
    \end{pgfonlayer}
    
   
\end{tikzpicture}
\end{scaletikzpicturetowidth}
\caption{A sequence of social sensors perform Bayesian social learning to estimate the underlying state $x$, and take a decision $a_k$ after myopically optimizing a reward function. The fusion center provides incentives $\cntrl_k \in [0,1]$ at each time $k$ (or at each sensor $k$) and fuses the information gathered in a Bayesian way. Each incentive $p_k$ is computed as a function $\mu$ of the posterior probability mass function (public belief) of the state $\pi_{k-1}$ at time $k-1$. The public belief $\pi_{k-1}$ is computed from the decisions of the first~$k-1$~sensors. The decision $a_k$ of social sensor $k$ depends on the incentive $p_k$, the public belief $\pi_{k-1}$, and the private observation $y_k$ of the state $x$. }
   \label{fig:SMIP}
   \end{figure}
We consider the setup illustrated in Fig.~\ref{fig:SMIP}. The fusion center controls the incentives given to the social sensors, and the social sensors share their decisions (quantized information on the underlying state) with the fusion center. Sec.\ref{sec:SLMIP} describes the controlled fusion social learning model that governs the manner in which the social sensors learn from each other, and how this behavior is influenced by the fusion center. Sec.\ref{sec:CIFO} formulates the objective of the fusion center that captures the trade-off between the cost of information acquisition from the social sensors versus the usefulness of the information measured.

\subsection{Controlled Fusion Social Learning Model} \label{sec:SLMIP}
In this subsection, we will model the following: (i) dynamics of the social sensors; (ii) the information fusion cost for the fusion center that models the trade-off between incentives and the reduction in uncertainty in the state estimate. We will also characterize the evolution of the posterior probability mass function of the state, and how the fusion center can make use of the available information to provide the incentives to the social sensors.

 Let $k = 1,2,\cdots$ denote the discrete time instants. It is assumed that each sensor decides once in a predetermined sequential order indexed by $k$. Let $x_0 (= x) \in \mathcal{X} = \{1,2\}$ denote the state of nature, and is assumed to be a {\em random variable}{\footnote{Sec.\ref{sec:IIFDS} discusses the estimation problem when the state is changing according to a Markov chain.} chosen at $k=0$. 
 %
 Let the probability mass function of the state $x$ at time $k-1$ be denoted as 
 \begin{equation} \label{eq:PubBel}
 \pi_{k-1}(i) = \Prb(x=i|a_{1},\hdots,a_{k-1}).
 \end{equation} 
 The state estimate (\ref{eq:PubBel}) is computed from the decisions of the social sensors $a_{1},\hdots,a_{k-1}$ and is termed as the {\em public belief}. Let the initial estimate be denoted as $\pi_{0} = (\pi_{0}(i),i\in \mathcal{X})$, where $\pi_{0}(i) = \Prb(x=i)$.  Let the belief space, i.e, the set of distributions $\pi$ over the state be denoted as $$\Pi(2){\overset{\Delta}{=}}\lbrace \pi \in \mathbb{R}^{2} : \pi(1)+\pi(2) = 1, 0 \leq \pi(i) \leq 1 ~\text{for} ~ i \in \{1,2 \}\}. $$

{\textbf{Social Sensor Dynamics:}} 
A social sensor, unlike a physical sensor, has its own dynamics since it learns from previous actions of other social sensors. It receives an observation on the underlying state, computes an estimate (private belief) using the information revealed by other sensors (their decisions), and takes an action to myopically maximize a reward function. This action/ decision is a quantization of the (private) belief, and is shared with the fusion center and other sensors. 
\begin{itemize}
\item[1.)]~\textit{{Social Sensor's Private Observation}}:  Each social sensor $k$'s obtains a noisy $y_{k} \in \mathcal{Y} = \{1,2\}$ of the underlying state~$x$ with the observation likelihood distribution:
\begin{equation}\label{eq:obs_m}
B_{ij} = \Prb(y_{k}=j|x=i).
\end{equation} 
The (discrete) observation likelihood distribution models the (limited) information gathering capabilities of the sensor. \\
\item[2.)]~\textit{{Social Learning and Private Belief update}}: Sensor $k$ updates its private belief $\eta_{y_k}$ by fusing observation $y_{k}$ and the prior public belief $\pi_{k-1}$, via the following classical Bayesian update
\begin{equation} \label{eq:PBU}
\eta_{y_k} = \frac{B_{y_{k}}\pi_{k-1}}{\textbf{1}'B_{y_{k}}\pi_{k-1}}
\end{equation}
where $B_{y_k}$ denotes the diagonal matrix with $[\Prb(y_{k}|x=1),~\Prb(y_{k}|x=2)]$ along the diagonal and $\textbf{1}^\prime$ denotes the $2$-dimensional row vector of ones. \\
\item[3.)]~\textit{{Social Sensor's Action}}: Sensor $k$ executes an action $a_{k}\in\mathcal{A}=\lbrace 1,2 \rbrace$ myopically to maximize a reward{\footnote{Each sensor being an expected (and myopic) reward maximizer is rational~\cite{Cha04}. This assumption implies that the social sensors have no altruistic concerns.}} function.  
The decision $a_k$ of social sensor $k$ is given by:
\begin{equation} \label{eq:AcS}
{\hspace{-0cm}}a_{k} =  {\underset{a \in \mathcal{A}}{\text{arg~max}}} ~r_a^{\prime}  \eta_{y_k}. 
\end{equation}
 Here $ r_a =[r(1,a),~r(2,a)]$, with $r_a^{\prime}$ denoting the transpose of the reward vector. We consider
\begin{align} \label{eq:Gam_xa}
r(1,a) &= \delta_a p_k + \Gamma_{1a},~~ r(2,a) = \delta_a p_k + \Gamma_{2a},~~ \text{with} \nonumber \\
~~\Gamma_{xa} &= - \alpha_a \Indc(a \neq x) - \gacq_a.
\end{align}
 Here $\delta_a \in [0,1]$,  $\alpha_a, \gacq_a \in \mathbb{R}$ are the given parameters of the model and $\Indc$ denotes the indicator function. For an action $a \in \mathcal{A}$ of the social sensor, $\delta_a p$ indicates the effective incentive received by the social sensor; $\gamma_a$  denotes the cost of taking the action; and $\alpha_a$ denotes the mis-representation or distortion weight~\cite{OLMRR17}. Appendix~\ref{App:Dis} provides a detailed discussion of the reward function, including the case where the reward is an explicit function of the observation.
 \end{itemize}
\textit{Tie-breaking rule}: When $r_a^{\prime}  \eta_{y_k} = r_{\bar{a}}^{\prime}  \eta_{y_k}, \forall \bar{a} \in \mathcal{A} / \{a\}  $, $a_k \sim \text{Uniform}(\mathcal{A})$, i.e, an action from the set $\mathcal{A}$ is chosen with probability $\frac{1}{|\mathcal{A}|}$, where $|\mathcal{A}|$ denotes the cardinality of set~$\mathcal{A}$. The uniform sampling tie-breaking rule ensures that the public belief (\ref{eq:PubBel}) is still a martingale. This is required in the proof of Theorem~\ref{thm:SMP}. 

{\textbf{Public Belief Dynamics:}}
The fusion center shares sensor $k$'s decision with the social sensors and the public belief (\ref{eq:PubBel}) is updated (by the fusion center and subsequent sensors) according to the social learning Bayesian filter (see \cite{Kri16, KB16}) as follows:
\begin{equation} \label{eq:SLF}
\pi_{k} = T^{\pi} (\pi_{k-1},a_k)  = \frac{R_{a_{k}}^{\pi_{k-1}}\pi_{k-1}}{\textbf{1}'R_{a_{k}}^{\pi_{k-1}}\pi_{k-1}}.
\end{equation}

Here, $R_{a_{k}}^{\pi_{k-1}} = \text{diag}(\Prb(a_{k}|x=i,\pi_{k-1}),i \in \mathcal{X})$ is the decision or action likelihood matrix (compare with the observation likelihood matrix $B$ in (\ref{eq:obs_m})), where 
\begin{align} \label{eq:ICH}
{\hspace{-0.0cm}}R^{\pi}_{ia} = \Prb(a_{k}|x=i,\pi_{k-1}) &= {\underset{y \in \mathcal{Y}}{\sum}} \Prb(a_{k}|y,\pi_{k-1})\Prb(y|x=i), ~~ \nonumber \\ \Prb(a_{k}|y,\pi_{k-1}) &= \left\{ \begin{array}{ll}
         1 & \mbox{if $ a_{k} =   {\underset{a \in \mathcal{A}}{\text{arg~max}}} ~r_a^{\prime}  \eta_{y_{k}}$} ; \\
         0 & \mbox{$\text{otherwise}$}.\end{array} \right.
\end{align}  
Note that $\pi_k \in \Pi(2)$. 
\begin{remark} [{\bf Information Cascade}]
Note that the (decision) likelihood probability (\ref{eq:ICH}) is an explicit function of the prior (public belief) $\pi_{k-1}$. This is unlike a standard Bayesian update (like (\ref{eq:PBU})), where the likelihood is independent of the prior. This unusual update of the social learning filter leads to herding behavior: In (\ref{eq:ICH}), if the action becomes independent of the observation, $R^{\pi}_{ia}  = 1~\text{or}~0$. This in turn leads to information cascade, social learning stops as the public belief is frozen, as can be seen from (\ref{eq:SLF}). It can be shown that (Theorem~$5.3.1$,~\cite{Kri16}) social learning stops in finite time.
\end{remark}
{\textbf{Fusion Center Dynamics:}} 
\begin{itemize}
\item[1.)] {\textit{Information Fusion cost:}} 
The fusion center minimizes the following cost of information fusion~$c(p_k)$, with 
\begin{equation} \label{eq:cst_es}
c(p_k) =  p_k  - \Phi_s (k) \Indc(a_k=y_k | \pi_{k-1}).
\end{equation}
Here $\Indc$ denotes the indicator function. The cost function should model the trade-off between incentives and truthful information disclosure{\footnote{Acting according to self valuations $(a = y)$ is in line with truthful information reporting in Peer Prediction literature; see~\cite{MRZ05}. We show in Sec.\ref{sec:CIF} that $a = y$ corresponds to informative decisions. Here informativeness is in the sense of Blackwell \cite{Kri16}; see also Footnote~$14$.}}. One possible{\footnote{In Sec.\ref{subsec:MTP}, we consider the information fusion cost that additionally has entropy of the state estimate.}} cost function is (\ref{eq:cst_es}). 
The information from different sensors is allowed to be weighed differently using $\Phi_s(k) \in (0,1)$. Here the subscript $s$ is used to denote the cost when only social learning is considered (see Sec.\ref{subsec:MTP} for the case when entropy cost, in addition to the effect of social learning, is considered). For simplicity, we assume the weights to be same for all sensors; i.e $\Phi_s(k) = \phi_s,~\forall~k$. Appendix~\ref{App:Dis} provides a motivation of the information fusion cost using well studied models in economics \cite{Ott96,Cha04,BOOV08}.
\item[2.)]{\textit{Information Fusion Incentive:}} 
The fusion center incentivizes/compensates the social sensors for providing information about the underlying state. The fusion center dynamically adapts these incentives over time as the sensors perform social learning: each sensor will have a different state estimate.  Let $\mathcal{F}_k$ denote the history of past incentives and decisions $ \{\pi_0,p_1,a_1,\cdots,p_{k-1},a_k \}$ recorded by the fusion center and the social sensors. More technically, 
\begin{equation} \label{eq:SgF}
{\hspace{-0cm}}\mathcal{F}_k := \sigma-\text{algebra generated by}~(\pi_0,a_1,\ldots,a_k,p_1,\dots,p_{k-1}).
\end{equation}
The fusion center chooses the incentive as $p_{k+1} \in \mu_{k}(\mathcal{F}_k)$ for the sensor $k+1$ to provide information about its state via social learning. Here $\mu_k$ denotes a policy that associates the history $\mathcal{F}_k$ with an incentive $p_{k+1}$. Since $\mathcal{F}_k$ is increasing with time $k$ (filtration), to implement a controller, it is useful to
obtain a sufficient statistic that does not grow in dimension. The public belief $\pi_{k}$ computed via the social learning filter (\ref{eq:SLF}) forms a sufficient statistic\footnote{See Sec.\ref{sec:SBSS} for justification.} for $\mathcal{F}_k$ and the incentive offered to social sensor $k+1$ is given as
\begin{equation} \label{eq:Pol_tim}
p_{k+1} = \mu_k(\pi_{k})~\in [0,1].
\end{equation}
The incentive is normalized to $[0,1]$ without loss of generality. 
\end{itemize}
\subsection{Controlled Information Fusion Objective} \label{sec:CIFO}
Given the setup in Sec.\ref{sec:SLMIP}, the aim of the fusion center is to estimate the state~$x_0(=x)$ by minimizing the cost of information acquisition ($p$). As discussed in (\ref{eq:SLF}), the fusion center performs Bayesian fusion of the information revealed by the social sensors. \\
Let $\bar{\boldsymbol{\mu} }= (\mu_0, \mu_1, \cdots)$ denote the sequence of policies employed by the fusion center at times $k = 0,1, \cdots$.
For each initial distribution $\pi_0$, the following cost is associated for the fusion center:
\begin{equation} \label{eq:RM}
J_{\bar{\boldsymbol{\mu} }}(\pi) = \E_{\bar{\boldsymbol{\mu} } } \{ \sum_{k=0}^{\infty} \rho^k c_{\mu_k}(p_k) | \pi_0 = \pi \}.
\end{equation}
Here $p_k$ denotes the incentive, $\rho \in [0,1)$ denotes an economic discount factor, $\mu_k$ denotes the decision policy (\ref{eq:Pol_tim}) for the fusion center that maps the public belief $\pi_k$ to an incentive $p_{k+1} \in [0,1]$, $c_{\mu_k}(p_k)$ denotes the cost of information fusion incurred at time $k$, and $\mathbb{E}_{\bar{\boldsymbol{\mu} }} $ denotes the expectation conditioned on the policy sequence $\bar{\boldsymbol{\mu} } $.

The policy sequence $\bar{\boldsymbol{\mu} }$ can be restricted to the class of stationary (time invariant) policies $\boldsymbol{\mu} = (\mu, \mu, \cdots)$ for the infinite horizon discounted cost objective; see~\cite{Kri16}. The fusion center aims to find the optimal stationary policy $\mu^*$ such that
\begin{equation} \label{eq:Opt_J}
J_{\mu^*}(\pi_0) = {\text{inf}}_{\mu \in \boldsymbol{\mu}} J_{\mu}(\pi_0)
\end{equation}
where $\boldsymbol{\mu}$ denotes the class of stationary policies.

{\bf Summary}: (\ref{eq:SLF}) are the dynamics and (\ref{eq:RM}) is the optimization objective for the controlled information fusion problem considered in this paper. The model parameters are the sensors' observation matrix $B$ in (\ref{eq:obs_m}) and the reward  $r_a$ in (\ref{eq:Gam_xa}). 

\subsection{Example: Social Media Review Platform} \label{subsec:Mot_Ex}
We briefly motivate above set-up using an application on online social media review platforms like Amazon or Airbnb. The state of a product $ x \in \{ 1(\text{Bad quality}), 2 (\text{Good quality}) \}$,  the observation $ y \in \{ 1(\text{Bad exp.}), 2 (\text{Good exp.}) \}$, and the customers' decision $a \in \lbrace 1(\text{Neg. Review}),2 (\text{Pos. Review})\rbrace$. When the customer writes a good/ bad review when it has a good/ bad  experience, the review is honest. Here it is assumed that each customer leaves a review, however, the nature of review depends on the optimization (\ref{eq:AcS}). 
The information fusion objective is to estimate the product or service quality and, Amazon or AirBnb want to maximize the number of customers that report honest experiences. This informative feedback from the social sensors can be used by the retailers to improve the quality, and it will also benefit the future customers in that they are well informed before making a decision. In this sense, the objective (\ref{eq:RM}) improves the overall welfare. 

\section{Structure of Optimal Incentive Policies} \label{sec:SOPP}
This section has three results. Sec.\ref{subsec:DPF} formulates solving for the optimal incentive policy (\ref{eq:Opt_J}) as a stochastic dynamic programming problem. Sec.\ref{subsec:SOIR} provides sufficient conditions on the model parameters $(B,r_a)$ under which the optimal incentive policy for the fusion center can be completely specified as a threshold policy. 
Sec.\ref{sec:POPS} provides a sample path characterization of the optimal incentive sequence that results from fusion center employing the optimal threshold policy.  

\subsection{Dynamic Programming Formulation} \label{subsec:DPF}
The optimal incentive policy $\mu^*$ in (\ref{eq:Opt_J}) and the corresponding optimal cost (value function) $V(\pi)$ satisfy the Bellman's stochastic dynamic programming equation \cite{Kri16}:
\begin{align} \label{eq:RMVP}
Q(\pi,\cntrl) &= c(\cntrl) + \rho \sum_{a \in \mathcal{A}} V(T^{\pi} (\pi,a)) \sigma(\pi,a), \nonumber \\ 
V(\pi) &= \min_{\cntrl \in [0,1]} Q(\pi,\cntrl), ~~ J_{\mu^*}(\pi_0) = V(\pi_0),~~\text{and}~~\\
~~\mu^*(\pi) &= {\underset{\cntrl \in [0,1]}{\text{arg~min}}}~Q(\pi,\cntrl). 
\end{align}
where  $T^{\pi} (\pi,a)$ is defined in (\ref{eq:SLF}) and $\sigma(\pi,a) = {\textbf{1}'R_{a}^{\pi}\pi}$, and $c(p)$ is the information fusion cost defined in (\ref{eq:cst_es}). \\
\textit{\underline{Discussion}}: Even though Bellman's equation~(\ref{eq:RMVP}) specifies the optimal policy, it has two problems: \\
 (i)~The state (belief) space $\Pi(2)$ is an uncountable set. Hence the dynamic programming equation~(\ref{eq:RMVP}) does not translate into practical solution methodologies, as the optimal cost $V(\pi)$ needs to be evaluated at each $\pi \in \Pi(2)$. \\
(ii)~The action (incentive) space for the information fusion center $p \in [0,1]$ is a continuum. It is well known \cite{Kri16} that even for a finite action case, computing the optimal policies is a computationally intractable PSPACE hard problem. 

\subsection{Structure of the Optimal Incentive Policy} \label{subsec:SOIR}
We wish to determine conditions under which the optimal incentive policy has the following intuitive threshold structure: don't incentivize if the estimate $\pi < \pi^*$, and incentivize using an exactly specified incentive function otherwise. Some of the advantages of the threshold policy are: (i) To compute the threshold policy (as in Fig.\ref{subfig1:STMT}), one only needs to compute the single belief $\pi^*$; whereas a general policy (as in Fig.\ref{subfig2:STMT}) requires PSPACE hard dynamic programming recursion offline. (ii)~To implement a controller with a threshold policy, one only needs to encode~$\pi^*$ and the incentive function, so its practically useful. \\ 
{\bf Incentive Function}: For future reference, we define the incentive function of the fusion center $\Delta(\eta_{y}) \in [0,1]$ as
\begin{equation}  \label{eq:Delt}
\Delta(\eta_{y}) = [l_1 ~ -l_2] \frac{B_{y}\pi}{\textbf{1}'B_{y}\pi} + l_3
\end{equation}
where $\eta_{y}$ is the private belief update (\ref{eq:PBU}) with $\pi_k$ replaced by $\pi$, $$l_1 = \frac{\alpha_2} {\delta_2 - \delta_1}, ~l_2 = \frac{\alpha_1} {\delta_2 - \delta_1},~l_3 = \frac{\gacq_2 - \gacq_1}{\delta_2 - \delta_1}.$$ 
The incentive function (\ref{eq:Delt}) naturally arises{\footnote{See also the proof of Theorem~\ref{lem:lmts}.}} by reformulating~(\ref{eq:AcS}). A set of parameters in the incentive function that ensure $\Delta(\eta_{y}) \in [0,1]$ are $l_1>0$, $l_2>0$ and $l_3>0$. A \textit{sufficient condition} is that $\alpha_1 > \alpha_2 $, $\delta_2> \delta_1$ and $\gacq_2 > \gacq_1$.  For other forms of reward functions (see Appendix~\ref{App:Dis}), the expression for $\Delta(\eta_{y}) \in [0,1]$ and the conditions on the model parameters are suitably derived.\\
\textbf{\underline{Model Assumptions}:}
We now give sufficient conditions under which the optimal incentive policy (\ref{eq:RMVP}) has a threshold structure.
\begin{compactenum}
\item[(A1)] The observation distribution $B_{xy} = \mathbb{P}(y|x)$ is TP2 (totally positive of order 2), i.e, the determinant of the matrix $B$ is non-negative.
\item[(A2)] The reward vector $r_a$ is supermodular, i.e, $r(1,1) > r(2,1)$ and $r(2,2) > r(1,2)$ for every $p \in [0,1]$.                                                                          
\end{compactenum}
%
 (A1)~is an assumption on the underlying stochastic model, and enables the comparison of the posteriors. The observation distribution being TP2 \cite{Kri16} implies that in higher states, the probability of receiving higher observations is higher than in lower states.   \\
 (A2)~is required for the problem to be non-trivial. If it does not hold and $r(i,1) > r(i,2)$ for $i = 1,2$, then $a=1$ always dominates $a=2$; the sensors provide no useful information. (see Sec.\ref{subsec:CFNB} for assumptions in non-binary environments)
\subsection*{\underline{Main Result: (Optimality of Threshold Incentive Policy)}}
Theorem~\ref{thm:OPRM} below is our first main result. It provides a closed form expression for the optimal policy $\mu^*(\pi)$ of the controlled information fusion problem: the optimal policy has threshold structure (as illustrated in Fig.\ref{subfig1:STMT}). The choice over a continuum of actions is reduced to a choice between two exactly specified incentive policies. The optimal policy is not unique\footnote{See the proof of Theorem~\ref{thm:OPRM}.}. There exists a version of the optimal policy having the structure as in Theorem~\ref{thm:OPRM}.
%
 %
\begin{theorem} \label{thm:OPRM}
Under (A1) and (A2), the optimal incentive policy defined in (\ref{eq:Opt_J}) is given explicitly as:
{\hspace{-1cm}} \begin{equation} \label{eq:OPESM}
 \mu^*(\pi) = \left\{ \begin{array}{lll}
         0 & \mbox{if $\pi(2) \in [0,\pi_s^*(2))$};\\
       \Delta(\eta_{y=2}) & \mbox{if $\pi(2) \in [\pi_s^*(2), 1]$}.\end{array} \right. 
        \end{equation}   
  Here the threshold state $\pi_s^*(2) \in (0,1)$ depends on the choice of $\phi_s \in (0,1)$ defined in (\ref{eq:cst_es}), and the parameters in the incentive function $\Delta(\eta_{y=2})$ defined in (\ref{eq:Delt}).    
\end{theorem}
%
\subsubsection*{\underline{Discussion}}\
The proof is given in the Appendix and involves establishing the following:  (i) show that due to the structure of the social learning filter in (\ref{eq:SLF}), the choice of incentives reduces from a continuum $[0,1]$ to a finite number at every belief; (ii) show that the incentive function $\Delta(\eta_y)$ is decreasing in $\pi$, hence the value function is monotone, and there exists a threshold~$\pi^*$. According to Theorem~\ref{thm:OPRM}, computing the optimal incentive policy is equivalent to finding the belief $\pi_s^*(2)$, below which it is optimal not to provide any incentive $p=0$; and above which it is optimal to incentivize  using $\Delta(\eta_{y=2})$ at every belief, to minimize the cost (see Fig.\ref{subfig1:STMT}). Therefore, the controlled information fusion problem reduces to a finite dimensional optimization problem of finding a threshold state $\pi^*$. Theorem~\ref{thm:OPRM} provides a closed form expression for the optimal policy of the controlled information fusion problem: the choice over a continuum of actions is reduced to a choice between two exactly specified policies: $\mu(\pi) = 0,~\forall~\pi$ and $\mu(\pi) = \Delta(\eta_{y=2}),~\forall~\pi$. 

The practical usefulness of Theorem~\ref{thm:OPRM} stems from the following: 
(i)~the search space of decision policies $\mu$ reduces from an infinite class of functions (over $\Pi(2)$) to those that switch once between the specified policies; (ii) at each instant (or belief) the fusion center only needs to decide between 
$p = \Delta(\eta_{y=2})$ and $p=0$; (iii) the region in the belief space $\Pi(2)$ where it is optimal to incentivize using $\Delta(\eta_{y=2})$ is connected and convex (compare Fig.\ref{subfig1:STMT} versus Fig.\ref{subfig2:STMT}). 
%
\subsection{Sub-martingale Property of Optimal Incentive Sequence} \label{sec:POPS}

%
Theorem~\ref{thm:OPRM} characterized the structure of the optimal incentive policy for controlled information fusion. A natural question is: How does the actual sample path of the optimal incentive sequence behave? Theorem~\ref{thm:SMP} below gives a sample path characterization of optimal incentive policy implemented by the fusion center. It is shown that when the fusion center aims to minimize the expected payout for gathering truthful information to reduce the uncertainty in the Bayesian state estimate, the incentive sequence is a sub-martingale{\footnote{See Appendix for definition.}}; i.e, it increases on average{\footnote{Here average is over different iterations of the estimation process. For example, each round of labelling/classification in Crowdsourcing can be seen as one iteration.}} over time. 
\begin{theorem} \label{thm:SMP}
Consider the information fusion problem with optimal policy $\mu^*(\pi)$ in (\ref{eq:OPESM}). Under (A1), the optimal incentive sequence~$\cntrl_k = \mu^*(\pi_{k-1})$ is a sub-martingale.
\end{theorem}
%
%
%
%
\subsubsection*{\underline{Discussion}}\
The proof is given in the Appendix and involves establishing the following: (i) the incentive function $\Delta(\eta_{y=2})$ is convex in $\pi$; (ii) the optimal incentive policy in (\ref{eq:OPESM}) is such that the incentives are increasing on average over time, using the notion of predictable sequences \cite{Dur10}. Typically in stochastic control problems, it is difficult to characterize the optimal control sequence; one can only characterize the optimal control policy. Theorem~\ref{thm:SMP} is interesting because we can characterize the optimal sequence of incentives as a sub-martingale. According to Theorem~\ref{thm:SMP}, the optimal incentive policy of the fusion center is such that the sample path of the incentive sequence displays an increasing trend, i.e, the incentives increase on average over time. \\
The usefulness of Theorem~\ref{thm:SMP} stems from the following: (i) it gives a sample path characterization of the optimal incentive policy implemented by the fusion center; (ii) the sub-martingale property assures that the average incentives should always increase over time. This is useful in assessing the reliability of the fusion center.\\
The increase in incentives over time can be attributed to the fact that the senors polled for information at a later instant have more accurate estimate of the state due to learning from predecessors, and hence require higher compensation to reveal the same. 
\section{Consistency of Controlled Information Fusion} \label{sec:CIF}
An elementary application of the martingale convergence theorem \cite{Dur10} shows that the social learning protocol (\ref{eq:SLF}) results in social sensors forming an information cascade; that is, after some time $n^*$, all sensors choose the same action and social learning stops (see Theorem~$5.3.1$,~\cite{Kri16}).
Therefore, the true state can never be estimated using social learning, indeed,  the belief will not converge to the true state asymptotically.

In this section, we show that by dynamically controlling the incentives over time, the fusion center can indeed learn the true state. However, this comes at the price of employing a sub-optimal incentive policy. We further provide uniform bounds on the additional cost incurred for consistency{\footnote{Let the true state be $x = \theta$. The pair ($\theta,\pi_k$) is consistent, if $\pi_k$ converges to a point mass at $\theta$ in probability.}}. When it is sufficient to know the state with a degree of confidence, policies that guarantee state estimation in finite time are discussed. We also provide uniform bounds on the budget saved as a result of estimating the state only upto a degree of confidence. 
\subsection{Controlled Information Fusion} 
Fig.\ref{fig:cf_f} shows the bi-directional interaction between the fusion center and the social sensor. The incentives chosen by the fusion center affects the reward function of the social sensors, and hence affects the decisions chosen. The decisions chosen in turn affect the estimate of the state (\ref{eq:PubBel}) for the fusion center as in (\ref{eq:SLF}). Recall that social learning terminates after a finite horizon (see remark on Information cascade after (\ref{eq:ICH})). Theorem~\ref{lem:lmts} below shows how to control the incentives to the social sensors to delay herding and information cascades, and hence estimate the state asymptotically. In particular, it is shown how the fusion center can \textit{control the incentives} such that the fusion of Bayesian estimates is consistent.
\begin{figure}[t]
	\vspace{-1cm}
	\begin{scaletikzpicturetowidth}{0.42\textwidth}
		\hspace{-0.15cm}  \begin{tikzpicture}[scale=\tikzscale]
		
		\node (wa) [sensor]  {Bayesian Filter};
		
		\path (wa.east)+(-4.5,2) node (vote) [wa] {Information \\ Fusion Center};
		
		\path (wa.west)+(4.5,2) node (asr1) [stat] {Social Sensor~$k$};
		
		\path [draw, thick, ->] (vote.east) -- node [midway, above] {Incentive~$p_k$} 
		(asr1.west);
		
		\path [draw, thick,->] (wa.west) -| node [midway,below] {Public Belief~$\pi_{k+1}$} 
		(vote.south);     
		
		\path [draw, thick, ->] (asr1.south) |- node [midway,below] {Decision~$a_{k}$} 
		(wa.east);   
		
		\path (asr1.north) +(0,1) node (asrs) {Private Belief~$\eta_{y_k}$};  
		
		\path [draw, dashed,thick,->] (asrs.south) -- node [midway,left] {} 
		(asr1.north); 
		
		\begin{pgfonlayer}{background}
		\path (asr1.west |- asr1.north)+(-6.5,1.5) node (a) {};
		\path (wa.north -| wa.north)+(+1,-1) node (b) {};
		\path (vote.east |- vote.north)+(+5.3,-3.2) node (c) {};
		
		\path[fill=brown!20,rounded corners, draw=black!50, dashed]
		(a) rectangle (c);           
		\path (asr1.north west)+(-3,3) node (a) {};
		
		\end{pgfonlayer}
		
		\end{tikzpicture}
	\end{scaletikzpicturetowidth}
	
	\caption{Bi-directional interaction between the information fusion center and the social sensor. The fusion center provides an incentive $p_k$ to the social sensor, which has a private belief $\eta_{y_k}$ after observation $y_k$. The social sensor takes a decision $a_k$ and this quantized information on the underlying state is used to update the public belief $\pi_{k+1}$ using a social learning Bayesian filter (\ref{eq:SLF}). The incentive $p_k$ at time $k$ directly modifies the reward function of the social sensor, and hence affects the state estimate $\pi_{k+1}$ at time $k+1$. }
	\label{fig:cf_f}
\end{figure}

We will express{\footnote{This is possible because of (A1) and (A2); see \cite{Kri12}.}} the belief space $\Pi(2)$ as a disjoint union of three connected regions to describe the sensors' decision dynamics as a function of the incentive $p$: a region $\mathcal{P}_1^\cntrl$ - where action $a=2$ is optimal; a region $\mathcal{P}_3^\cntrl$ - where action $a=1$ is optimal; a region $\mathcal{P}_2^\cntrl$ - where action $a=y$ is optimal. From (\ref{eq:AcS}), the decision of the social sensor depends on the private belief $\eta_y$ and the reward $r_a$ (defined in (\ref{eq:Gam_xa})). Therefore, define:
\begin{align} \label{eq:REG}
\mathcal{P}_1^\cntrl &= \{ \pi \in \Pi(2): (r_1 - r_2)^{\prime} \eta_{y=1}  \leq 0 \} \nonumber \\
\mathcal{P}_2^\cntrl &= \{  \pi \in \Pi(2): (r_1 - r_2)^{\prime} \eta_{y=1}  > 0  \cap (r_1 - r_2)^{\prime} \eta_{y=2}  \leq 0 \} \nonumber \\
\mathcal{P}_3^\cntrl &= \{ \pi \in \Pi(2): (r_1 - r_2)^{\prime} \eta_{y=2}  > 0\}
\end{align}
where $r_a$ for $a = \{1,2\}$ are the social sensors' rewards and $\mathcal{P}^\cntrl$ models the explicit dependence of the width of the regions on the incentive parameter $p$ through $r_a$, $\eta_{y=1}$ and $\eta_{y=2}$ denote the private belief updates after $y=1$ and $y=2$ respectively. The region $\mathcal{P}_1^\cntrl \cup \mathcal{P}_3^\cntrl$ is the \textit{herding} region and $\mathcal{P}_2^\cntrl$ is the \textit{social learning} region for any $p \in [0,1]$. 
\begin{theorem} \label{lem:lmts}
Under (A1) and (A2), the following relation holds between the incentive $p_{k}$ and the public belief $\pi_{k+1}$:
\begin{equation*} \label{eq:Prc}
\hspace{-1.5cm} \pi_{k+1} \in \left\{ \begin{array}{lll}
         \mathcal{P}_3^\cntrl & \mbox{iff $p_k  \in [0,\Delta(\eta_{y_k=2}))$};\\
       \mathcal{P}_2^\cntrl & \mbox{iff $p_k  \in [\Delta(\eta_{y_k=2}), \Delta(\eta_{y_k=1}))$}; \\
       \mathcal{P}_1^\cntrl & \mbox{iff $p_k  \in [\Delta(\eta_{y_k=1}), 1]$}.\end{array} \right. 
 \end{equation*}    
where the regions $\mathcal{P}_i^\cntrl$ for $i = 1,2,3$ are defined in (\ref{eq:REG}), and $\Delta(\eta_y)$ is as in (\ref{eq:Delt}). 
\end{theorem}
 \subsubsection*{\underline{Discussion}}\
The proof is given in the Appendix. According to Theorem~\ref{lem:lmts}, relation between the incentive $p_k$ at time $k$ and the state estimate (public belief~$\pi_{k}$) at the next instant $k+1$ is such that, when $p_k$ belongs to the intervals defined by the private beliefs (in the incentive function $\Delta(\eta_{y_k}))$, the widths of the herding and social learning regions change (see~Fig.\ref{fig:cif}) so that the public belief ($\pi_{k+1}$) belongs to the desired~$\mathcal{P}_i^\cntrl$. Fig.\ref{fig:cif} shows the variation of the width of the regions with respect to the incentive parameter~$p$. 
 Theorem~\ref{lem:lmts} characterizes the sensitivity of the regions $\mathcal{P}_1^\cntrl, \mathcal{P}_2^\cntrl, \mathcal{P}_3^\cntrl$ with respect to the incentive $p \in [0,1]$, and Corollary~\ref{cor:AL} below shows how to stop the information cascade so that social learning can proceed indefinitely so that the state estimate converges to the true state.  
 \begin{figure}[!h] 
\captionsetup{singlelinecheck=off}
\centering
{\hspace{0.2cm}}\includegraphics[scale=0.3]{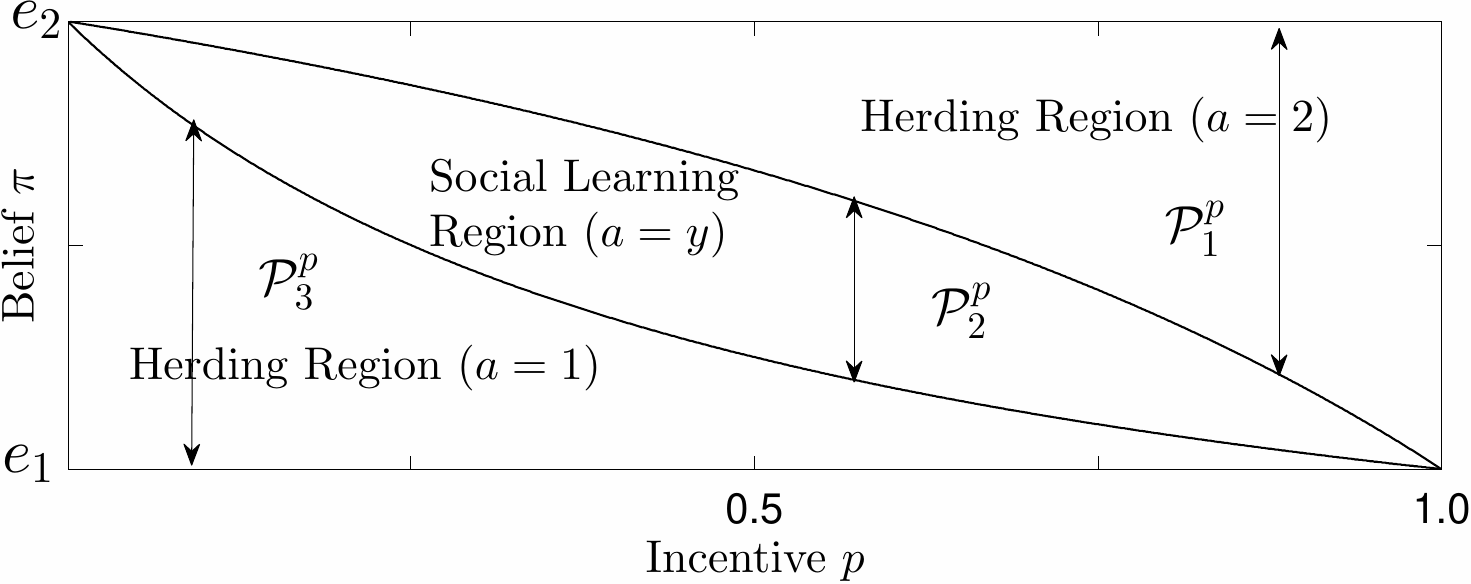}
\caption{Herding ($\mathcal{P}_1^\cntrl \cup \mathcal{P}_3^\cntrl$) and social learning ($\mathcal{P}_2^\cntrl$) regions with respect to the incentive parameter $p$. 
	It is seen that when the incentives are small (close to~$0$), the sensors herd on low quality actions ($a=1$); and when the incentives are high (close to~$1$), the sensors herd on high quality actions ($a=2$); however, only the actions in the social learning region are informative or reflect the sensors' true valuation.}
\label{fig:cif}
\end{figure}  
%
\begin{corollary} \label{cor:AL}
Let $p_k = \Delta(\eta_{y_k=2})$ for $k = 1,2,\hdots$. The fusion of Bayesian estimates is consistent, i.e, the fusion center learns the true state asymptotically.
\end{corollary}
\subsubsection*{\underline{Discussion}}\
We know that the fusion center can force the state estimates to be in the social learning region by choosing incentives in the range $p \in [\Delta(\eta_{y=2}), \Delta(\eta_{y=1}))$, see Fig.\ref{fig:cif}. From (\ref{eq:REG}), Lemma~\ref{lem:IncPB} and Theorem~\ref{thm:VFRM} in the Appendix, the social sensors' decision likelihood matrices $R^{\pi}_a$ (as in (\ref{eq:SLF})) in regions $\mathcal{P}_1^\cntrl, \mathcal{P}_2^\cntrl, \text{and}~ \mathcal{P}_3^\cntrl$ for any $p \in [0,1]$ are
\[
\begin{bmatrix}
0 & 1\\
0 & 1
\end{bmatrix},
\begin{bmatrix}
B_{11} & B_{12}\\
B_{21} & B_{22}
\end{bmatrix},
~\text{and}~
\begin{bmatrix}
1 & 0\\
1 & 0
\end{bmatrix}
\]  
respectively.  In the herding region $\mathcal{P}_1^\cntrl \cup \mathcal{P}_3^\cntrl$, the decision of the social sensor is independent of the public belief and the public belief (\ref{eq:SLF}) is frozen. In the social learning region $\mathcal{P}_2^\cntrl$, the sensors take informative{{\footnote{Informativeness is in the sense of Blackwell; see \cite{Kri16}. For any two observation matrices $B_1$ and $B_2$, $B_1$ is more informative than $B_2$ in the Blackwell sense ($B_1 \succ_B B_2$) if $B_2 = B_1\Gamma$, for any stochastic matrix $\Gamma$. When the sensors act according to their observations,  $\pi \in \mathcal{P}_2^\cntrl$, and the decision likelihood matrix in (\ref{eq:SLF}) $R^\pi_S=B$; and when the sensors don't act according to the observations (they herd), $\pi \in \mathcal{P}_3^\cntrl$, the decision likelihood matrix $R^\pi_H=\begin{bmatrix}
1 & 0\\
1 & 0
\end{bmatrix}$. We have for $\Gamma = \begin{bmatrix}
1 & 0\\
1 & 0
\end{bmatrix},~$ $R^\pi_H = R^\pi_S \Gamma \Rightarrow R^\pi_S \succ_B R^\pi_H$.}} actions; i.e, each sensor acts according to its observation/valuation.  In the social learning region,  sensors take informative actions $a=y$; or $R^\pi_a = B$. The observations are conditionally independent given the true state. Therefore, by suitably controlling the incentives, the fusion center fuses information that is i.i.d on the true state. It is well known \cite{DF86,Van98} that fusion of Bayesian estimates is consistent (convergence in probability); i.e, for a point mass at the true state $\theta$ denoted as $g(\theta)$,
$$\lim_{k \rightarrow \infty} \Prb(|\pi_k - g(\theta)| > \epsilon) = 0~ \forall~\epsilon > 0.$$
In other words, the fusion center \textit{can} learn the true state asymptotically by choosing the incentives as $p_k = \Delta(\eta_{y_k=2})$ for $k = 1,2,\ldots$. 

\subsection{Cost of consistency for the fusion center} \label{subsec:CCFC}
When the incentive policy is the optimal threshold policy~(\ref{eq:OPESM}), the fusion of Bayesian estimates computed from the social sensors' decisions (\ref{eq:SLF}) is not consistent. This is because, the optimal incentive policy for the fusion center is such that below a certain threshold it is optimal to not incentivize (see Fig.\ref{subfig1:STMT}). From Theorem~\ref{lem:lmts}, when the fusion center stops incentivizing $p = \mu^*(\pi) = 0$, the public belief is in the herding region $\mathcal{P}^p_3$. In the herding region, social learning ceases and there is no improvement in uncertainty -- mean square error between the state estimate and the true parameter remains at a fixed non-zero value. 
If, however, the fusion center chooses a sub-optimal policy (\ref{eq:CP}), it will incur additional cost for the incentives; but the fusion of estimates computed from the social sensors' decisions (\ref{eq:SLF}) will be consistent (Corollary~\ref{cor:AL}). Theorem~\ref{thm:AHI} below provides uniform bounds on the additional cost incurred by the fusion center for employing a sub-optimal incentive policy that results in consistent information fusion. \\
Consider the objective function for the fusion center:
\begin{equation} \label{eq:WM}
W_{\mu_c}(\pi) = \E_{\mu_c} \{ \sum_{k=0}^{\infty} \rho^k c_{\mu_c}(p_k) | \pi_0 = \pi \}
\end{equation}
where $W_{\mu_c}(\pi)$ denotes the cost incurred by employing the sub-optimal policy (compare with (\ref{eq:OPESM}))
{\hspace{-1cm}} \begin{equation} \label{eq:CP}
\mu_c(\pi) = \left\{  \Delta(\eta_{y=2}) ~\forall ~\pi(2) \in [0, 1] \right \}. 
        \end{equation} 
\begin{theorem} \label{thm:AHI}
Let (A1) hold. The additional cost (on average) incurred by the fusion center for employing the sub-optimal policy $\mu_c(\pi)$ in (\ref{eq:CP}) instead of the optimal policy $\mu^*(\pi)$ in~(\ref{eq:OPESM}) is bounded as:
\begin{equation}
\sup_\pi |W_{\mu_c}(\pi) - J_{\mu^*}(\pi)| \leq  2 \frac{(1 - \phi_s)}{1 - \rho}
\end{equation}
where $J_{\mu^*}(\pi)$ is the optimal cost (\ref{eq:Opt_J}).
\end{theorem}
\subsubsection*{\underline{Discussion}}\
The proof is given in the Appendix and uses the fact that both $W_{\mu_{c}}(\pi)$ and $J_{\mu^*}(\pi)$ are decreasing in $\pi$ established using similar arguments as in Theorem~\ref{thm:OPRM}, and $W_{\mu_{c}}(\pi) - J_{\mu^*}(\pi) \geq 0 ~ \forall~\pi$. Theorem~\ref{thm:AHI} characterizes the trade-off between consistency and cost of information acquisition. It says that when the fusion center employs a sub-optimal policy, the average additional cost incurred is bounded above by the weight $\phi_s$ in the information fusion cost~(\ref{eq:cst_es}), discount factor~$\rho$ that captures the degree of impatience of the fusion center. 

The usefulness of Theorem~\ref{thm:AHI} stems from the following: (i)~It gives an upper bound on the additional discounted cost incurred when the fusion center chooses the incentives such that the fusion of Bayesian estimates computed as in (\ref{eq:SLF}) is consistent. (ii)~It helps in choosing the weight $\phi_s$ and the discount factor $\rho$ for the fusion center. \\  

\subsection{Finite time bounds for the fusion center}
In Sec.\ref{subsec:CCFC}, it was shown that by employing a sub-optimal policy the fusion center can estimate the true state asymptotically. However, it is often enough to know the state with a degree of confidence. In this section, we obtain uniform bounds on the budget saved by estimating the state upto a degree of confidence.

The degree of confidence characterizes regions in the belief space $\Pi(2)$ that can be used to estimate the states. For a degree of confidence $\vartheta \in (0,1)$, any belief in the confidence region $\pi(2) \in [0, \vartheta]$ is identified with state $x = 1$, and any belief in the confidence region $\pi(2) \in [1-\vartheta,1]$ is identified with state $x = 2$. For example, when the public belief (posterior) is such that $\pi(2) \in [0.9,1]$, then the fusion center is (atleast) $90\%$ confident that the state $x=2$, and if $\vartheta < 0.1$, the state is estimated as $x=2 $.  
For a degree of confidence $\vartheta \in (0,1)$, consider using the following policy 
\begin{equation} \label{eq:Fin_Pol}
\mu_f(\pi) = \left\{ \begin{array}{ll}
0 & \mbox{if $\pi(2) \in [0, \vartheta]$};\\
\Delta(\eta_{y_k}) & \mbox{if $\pi(2) \in (\vartheta, 1-\vartheta)$};\\
0 & \mbox{if $\pi(2) \in [1-\vartheta,1]$}.\end{array} \right. 
\end{equation} 
It can be shown using martingale convergence theorem \cite{Dur10} that when using the policy (\ref{eq:Fin_Pol}), the public belief hits one of the two confidence regions in finite time{\footnote{The arguments are similar to those used to establish information cascades occur in finite time in \cite{BHW92} and Theorem~$5.3.1$,~\cite{Kri16}.}}. \\
The following theorem provides a bound on the budget saved by employing the policy in (\ref{eq:Fin_Pol}) instead of the policy (\ref{eq:CP}). Let $\pi_\vartheta = \begin{bmatrix}
\vartheta \\
1 - \vartheta
\end{bmatrix}$ and $\eta_\vartheta = \frac{B_{y=2}\pi_\vartheta}{\boldsymbol{1}^\prime B_{y=2}\pi_\vartheta}$. 
\begin{theorem} \label{thm:AHF}
	Let (A1) hold. For a degree of confidence $\vartheta$, the budget saved by the fusion center by employing the policy $\mu_f(\pi)$ in (\ref{eq:Fin_Pol}) instead of the policy $\mu_c(\pi)$ in~(\ref{eq:CP}) is bounded as:
\begin{equation}
\begin{split}
    \sup_\pi |W_{\mu_c}(\pi) - W_{\mu_f}(\pi)| &\leq   2\frac{(1 - \phi_s)}{1 - \rho} + \frac{|\Delta(\eta_\vartheta) - \phi_s|}{1 - \rho}.
\end{split}
\end{equation}	
	where $\rho$ is the discount factor.
\end{theorem}

\subsubsection*{\underline{Discussion}}\
The proof follows using arguments similar to Theorem~$5$ in the paper. Theorem~\ref{thm:AHF} provides an uniform bound on the budget saved by employing the policy $\mu_f(\pi)$ in~(\ref{eq:Fin_Pol}) instead of $\mu_c(\pi)$ in (\ref{eq:CP}). A bound on the budget saved with respect to the optimal policy $\mu^*(\pi)$ can be obtained from Theorem~\ref{thm:AHF} and Theorem~$\ref{thm:AHI}$ using the triangle inequality of the norm,
\begin{equation*}
|J_{\mu^*}(\pi) - W_{\mu_f}(\pi)| \leq |W_{\mu_c}(\pi) - J_{\mu^*}(\pi) |  + |W_{\mu_c}(\pi) - W_{\mu_f}(\pi)|. 
\end{equation*}
In Theorem~\ref{thm:AHF}, the fact (Lemma~\ref{lem:D_dec}) that $\Delta(\eta_{y=2})$ is decreasing in $\pi$, and $|\varepsilon| \geq \varepsilon$ is utilized in deriving the bounds.

%
\section{Strategic behaviour in Social Sensors} \label{sec:SBSS}
The information fusion center polls the social sensors in a pre-determined order and they decide what information to reveal, i.e, it was assumed that the sensors do not hide their signals and are not strategic. However, the rewards can be suitably designed so that the sensors reveal information when polled.
In this section, we show how to design the reward functions to prevent the social sensors from being strategic. This implies that the social sensors  have no forward-looking tendencies and reward function of the social sensors has no externalities, and the public belief (\ref{eq:PubBel}) forms a sufficient statistic for the history of past actions and incentives. 

Under an additional minor restriction{\footnote{This is independent of the actual form of the rewards when the rewards in both states are non-zero.}} on the reward parameters, it is shown below that the social sensors have no incentive to delay or hide their signals. 

\textit{\underline{Social sensors do not display contrarian behavior}}: \\ The optimal policy for the fusion center dictates that it either incentivize or not incentivize, see Theorem~$\ref{thm:OPRM}$. When the fusion center is offering incentives ($\Delta(\eta_{y=2})$), from Theorem~$\ref{lem:lmts}$, it is seen that it is optimal for the social sensors to act according to their observations. As the social sensors are assumed to be Bayes rational, they have no incentive to deviate. When the fusion center is not incentivizing, the sensors always herd.

\textit{\underline{Social sensors are not strategic}}: Let $\mathcal{R}_H$ and $\mathcal{R}_S$ denote the regions where the fusion center does not incentivize ($\mu(\pi) = 0$) and incentivizes ($\mu(\pi) = \Delta(\eta_{y=2})$)  respectively.  A social sensor deciding at time $k$ considers the following scenarios:

\begin{itemize}
	\item[a.)] $\pi_{k} \in \mathcal{R}_S$ and $\pi_{k+1} \in \mathcal{R}_H$.  In other words if the sensor delays revealing information and the belief update after the next ($k+1$) sensors' decision belongs to the region where there is no incentivization. \\
	$p_{k+1} = 0$, so the social sensor $k$ would be better off revealing at time $k$.
	\item[b.)] $\pi_{k},~\pi_{k+1} \in \mathcal{R}_S$ and $\pi_{k+1}(2) < \pi_{k}(2)$.
	
	Consider the rewards for the social sensor from (\ref{eq:Gam_xa}), 
	\begin{align} \label{eq:RSS}
r_{1} &= [\delta_1 p + \Gamma_{11} ~~ \delta_1 p + \Gamma_{21}] \nonumber \\
r_{2} &= [\delta_2 p + \Gamma_{12} ~~ \delta_2 p + \Gamma_{22}] 
	\end{align}
Assume $\Gamma_{ij}>0$ for all $i,j$ without loss of generality. Note that the reward vector $r_a$ is also required to be super-modular for any $p$ , so $\Gamma_{11} > \Gamma_{21}$ and $\Gamma_{22} > \Gamma_{12}$.
	Let $T(\pi,y_k) = \frac{B_{y_k}\pi}{\boldsymbol{1}^\prime B_{y_k}\pi}$ denote the private belief of sensor $k$. There are two possible observations for the social sensor $k$, $y_k = 1,2$. We will establish the result for $y_k=1$, and the result follows immediately for $y_k=2$.\\
Let $\bar{r}_a = [\delta_a p_{k+1} + \Gamma_{1a} ~~ \delta_a p_{k+1} + \Gamma_{2a}]$. 
	\begin{proposition} \label{prop:FLA1}
	Let the observation of sensor $k$ be $y_k=1$. There is a discount factor $D \in (0,1]$ such that $r_1^\prime T(\pi_k,y_k=1) \geq D~\bar{r}_1^\prime T(\pi_{k+1},y_k=1) $.
	\end{proposition}
	 	 
	\textbf{Proof}: From the definition of First-order stochastic dominance and TP2 on $B$, we have the following{\footnote{Note that for any $a,b>0$, $a<b \Rightarrow a > (\frac{a}{b}-\epsilon)b$ for any $\epsilon > 0$.}}
	\begin{align*}
	r_1^\prime T(\pi_k,y_k=1) &\leq  \bar{r}_1^\prime T(\pi_{k+1},y_k=1) \\
	\therefore r_1^\prime T(\pi_k,y_k=1) &> D~\bar{r}_1^\prime T(\pi_{k+1},y_k=1), \\
	\text{where}~D &= \frac{r_1^\prime T(\pi_k,y_k=1)}{\bar{r}_1^\prime T(\pi_{k+1},y_k=1)} - \epsilon, \text{for}~\epsilon > 0.
	\end{align*}
	Considering the largest possible deviation{\footnote{Note that $\pi_{k},~\pi_{k+1} \in \mathcal{R}_S$. Clearly, this is included in $\pi_k(2), \pi_{k+1}(2) \in [0,1]$.}} $\pi_k(2) = 1$ and $\pi_{k+1}(2) = 0$, it is easily seen that the smallest value for $D = \frac{\Gamma_{21}}{\delta_1 + \Gamma_{11}} - \epsilon < 1$. \qed
	
	\textit{Discussion}: The social sensors are not more forward looking than~$D$ from Proposition~\ref{prop:FLA1}. By suitably choosing the reward parameters, we can obtain $D=1$. This implies that the social sensors have no incentive to deviate when $y_k=1$.
	\item[c.)] $\pi_{k},~\pi_{k+1} \in \mathcal{R}_S$ and $\pi_{k+1}(2) > \pi_{k}(2)$. By using similar arguments as in Proposition~\ref{prop:FLA1}, we obtain the discount factor $D = \frac{\Gamma_{12}}{\delta_2 + \Gamma_{22}} - \epsilon < 1$. By suitably choosing the reward parameters, we can obtain $D=1$. This implies that the social sensors have no incentive to deviate when $y_k=2$. Also, the result follows immediately for $y_k=1$.
\end{itemize}	
It was shown that when the reward parameters are chosen so that $D=1$, myopically maximizing the expected reward is a Markov perfect equilibrium.

\section{Controlled Information Fusion with Dynamic States} \label{sec:IIFDS}
So far, we considered the problem of incentivized information fusion for estimating the random variable $x \in \mathcal{X}$. In this section, we consider the information fusion to estimate the state of a Markov chain $x_k$ for $k = 0,1,2,\cdots$ with social sensors. The dynamic states might correspond to, for example, a change in the product/ service quality on AirBnb or Amazon.

Let the state $x_k$ evolve as a Markov chain on the space $\mathcal{X}$ with a transition probability matrix $P$ and an initial distribution $\pi_0$ in (\ref{eq:PubBel}). Below we briefly highlight the changes in the social learning model in Sec.\ref{sec:SLMIP} for the case of dynamic states. \\
The private belief update in (\ref{eq:PBU}) for the social sensors taking the possible state change into account is given as  \begin{equation} \label{eq:PBU_mc}
\eta_{y_k} = \frac{B_{y_{k}} P^{\prime} \pi_{k-1}}{\textbf{1}'B_{y_{k}} P^{\prime} \pi_{k-1}}
\end{equation}
The public belief update in (\ref{eq:SLF}) taking the possible state change into account is given as
\begin{equation} \label{eq:SLF_mc}
\pi_{k} = T^{\pi} (\pi_{k-1},a_k)  = \frac{R_{a_{k}}^{\pi_{k-1}} P^{\prime} \pi_{k-1}}{\textbf{1}'R_{a_{k}}^{\pi_{k-1}} P^{\prime} \pi_{k-1}}.
\end{equation}

The optimal incentive policy in case of a random variable $\mu^*(\pi)$ in Theorem~\ref{thm:OPRM} is near optimal for the case of dynamic states, when transitions out of the current state is allowed only with a small probability. 
This is shown in Theorem~\ref{thm:SETr} below. 

Let $\mu^*(\pi)$ denote the optimal policy for estimating/ localizing the random variable ($P = I$); and $\mu_\epsilon^*(\pi)$ denote the optimal policy for estimating/ tracking the state of a Markov chain with $P = \begin{bmatrix}
1 - \epsilon_{1} & \epsilon_{1} \\
\epsilon_{2} & 1 - \epsilon_{2}
\end{bmatrix}$, where~$\epsilon_{1},\epsilon_{2}>0$.

\begin{theorem} \label{thm:SETr}
	Let $\rho \in [0,1)$ denote the economic discount factor. Let $V_{\mu^*}(\pi)$ and $V_{\mu_\epsilon^*}(\pi)$ denote the optimal costs incurred by employing the optimal policy $\mu^*(\pi)$ and $\mu_\epsilon^*(\pi)$ respectively. The following holds:
	\begin{align}
 V_{\mu^*}(\pi) - V_{\mu_\epsilon^*}(\pi) &\leq \frac{2 \rho (1 - \phi_s) (\epsilon_{1} + \epsilon_{2})}{(1-\rho)^2} ~~\times \nonumber \\ &\max \{|B_{21} - B_{11}|, |B_{22} - B_{12}| \}.
	\end{align}
\end{theorem}

\subsubsection*{\underline{Discussion}} \ The proof follows from Theorem $2$ in \cite{RIMB09}. Theorem~\ref{thm:SETr} says that the policy $\mu^*(\pi)$ incurs a total cost $V_{\mu^*}(\pi)$ that is within  $O(\epsilon_{1}+\epsilon_{2})$ of the total cost $V_{\mu_\epsilon^*}(\pi)$. When $\epsilon_{1}, \epsilon_{2} << 1$, the policy $\mu^*(\pi)$ for the state localization problem ($P = I$) is near optimal for the state tracking problem~($P \neq I$).

Characterizing the nature of the optimal incentive sequence (as in Sec.\ref{sec:SOPP}) in case of a random variable relied on the crucial fact that the belief is a martingale unconditional on the state. However, when the states are changing, the public belief (\ref{eq:SLF_mc}) is not a martingale (see \cite{Cha04}). This implies that, even though, $\mu^*(\pi)$ is near optimal, the incentive sequence that results from the fusion center employing $\mu^*(\pi)$ need not show an increasing trend on average.

%
\section{Numerical Results} \label{sec:NuEx}
Sec.\ref{subsec:MTP} below illustrates a controlled information fusion with quadratic cost unlike~(\ref{eq:cst_es}). It is shown that a multi-threshold incentive policy is optimal for the fusion center.  Sec.\ref{subsec:SOMP} illustrates the sensitivity of the optimal threshold (\ref{eq:OPESM}) to the parameters $\phi_s$ (the weight in (\ref{eq:cst_es})) and $\rho$ (discount factor in the objective (\ref{eq:RM})) that are chosen by the fusion center.  Sec.\ref{subsec:SPOI} illustrates the relation between the information gathering capabilities of the sensor (observation matrix $B$ in (\ref{eq:obs_m})) and the average incentives provided by the fusion center. Sec.\ref{subsec:CFNB} discusses the formulation and a numerical simulation for the controlled information fusion in non-binary environments.

Bellman's equation (\ref{eq:RMVP}) is solved by discretizing the state space $\Pi(2)$.  The optimal incentive policy and the optimal cost for the fusion center are computed by constructing a uniform grid of $1000$ points for $\pi(2) \in [0,1]$ and then implementing the policy and value iteration algorithm~\cite{Kri16} for a duration of $N=100$.  
\begin{table} [h!]
\begin{center}
\begin{tabular}{ |c|c|c| } 
 \hline
 $\alpha_1 = 0.288$ & $\alpha_2 = 0.278$ & $\beta_1 = 0.11$ \\ \hline 
  $ \beta_2 = 0.1$ & $\gacq_1 = 0.1$ & $\gacq_2 = 0.414$ \\
 \hline
\end{tabular} 
\end{center} 
\caption{For $\delta_1 = 0.3, ~\delta_2 = 0.95$, the following parameters were obtained as a solution of $\Delta(e_1)=1$ and $\Delta(e_2)=0$ for the reward vector (\ref{eq:res_d_rwrd}) parameters with the observation matrix $B$=\usebox{\smlmat}. }
\label{eq:Para}
\end{table}
\begin{figure*}[t!]
    \begin{subfigure}[t]{0.48\textwidth}
        {\hspace{-0.2cm}}\includegraphics[width=6.5cm,height=2.5cm]{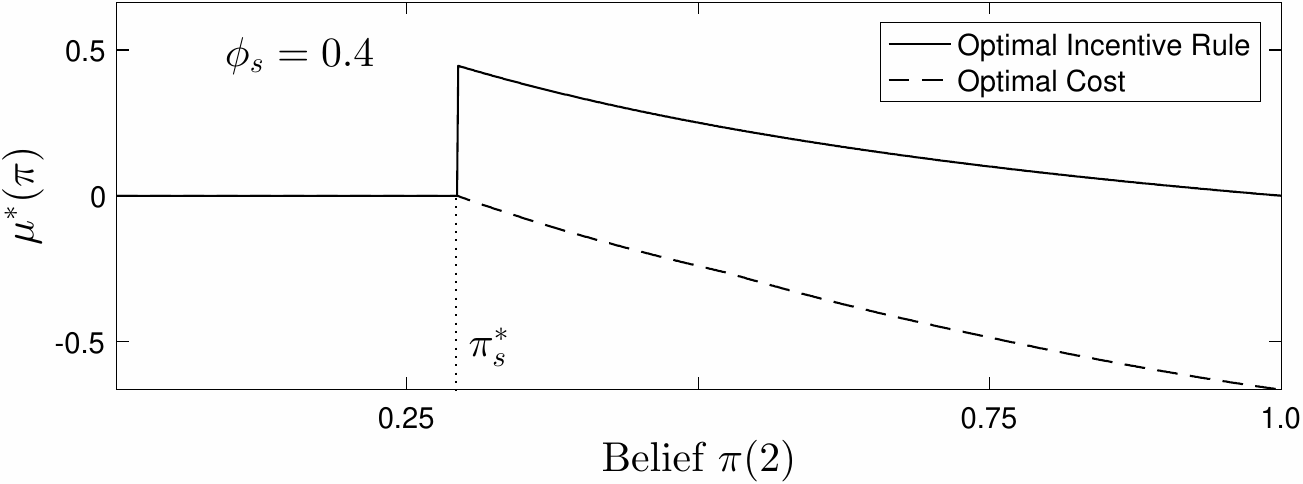}
        \caption{Optimal Policy for social learning weight $\phi_s = 0.4$.}
        \label{subfig1:phi1}
    \end{subfigure}%
    ~ 
    \begin{subfigure}[t]{0.48\textwidth}
       {\hspace{0cm}}\includegraphics[width=6.5cm,height=2.5cm]{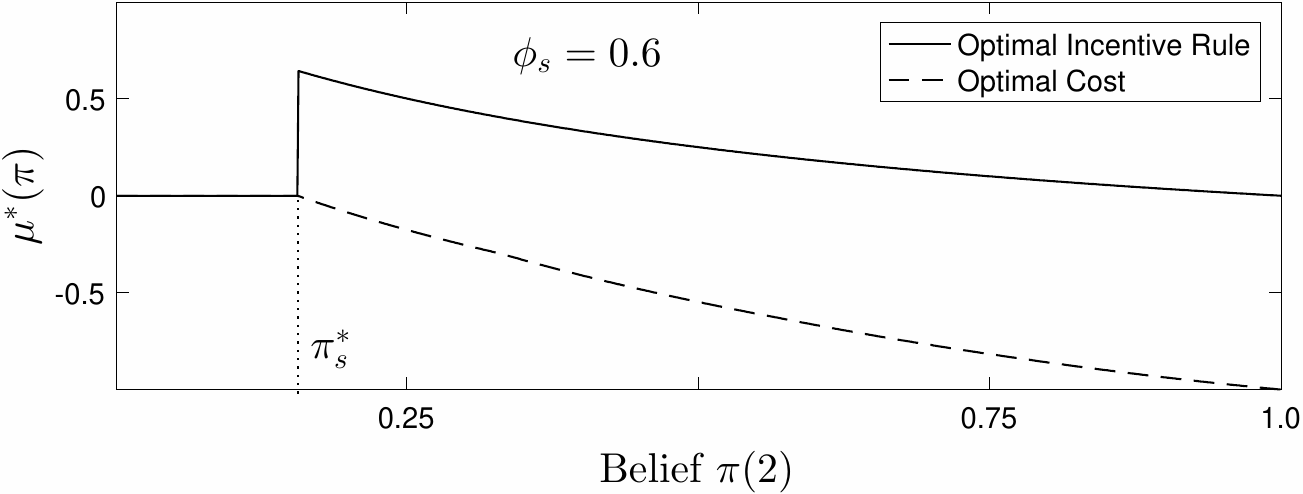}
        \caption{Optimal Policy for social learning weight $\phi_s = 0.6$.}
        \label{subfig2:phi2}
    \end{subfigure}
    \caption{Usefulness of information vs Incentivizing trade-off for the fusion center. It can be seen that $\pi^*(2)$ is decreasing with $\phi_s$ -- a higher weight will necessitate incentivizing sooner. According to Theorem~\ref{thm:AHI}, higher $\phi_s$ implies that the additional cost for employing a sub-optimal policy is smaller; in other words, $\pi_s^*$ is smaller. The parameters of the incentive function (\ref{eq:Delt}) are given in Table~\ref{eq:Para} and the discount factor $\rho = 0.4$. Here $\phi_s$ denotes the weight in the information fusion cost (\ref{eq:cst_es}). }
\label{fig:Tr_phi}
\end{figure*}
\begin{figure*}[t!]
    \begin{subfigure}[t]{0.48\textwidth}
        {\hspace{-0.2cm}}\includegraphics[width=6.5cm,height=2.5cm]{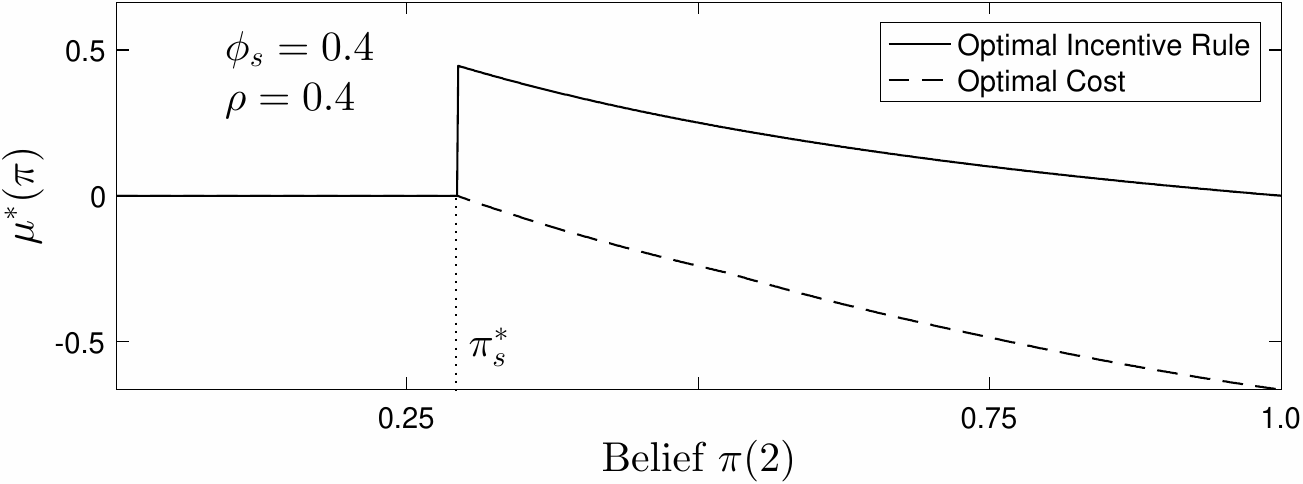}
        \caption{Optimal Policy for discount factor $\rho = 0.4$.}
        \label{subfig1:rho1}
    \end{subfigure}%
    ~ 
    \begin{subfigure}[t]{0.48\textwidth}
       {\hspace{0cm}}\includegraphics[width=6.5cm,height=2.5cm]{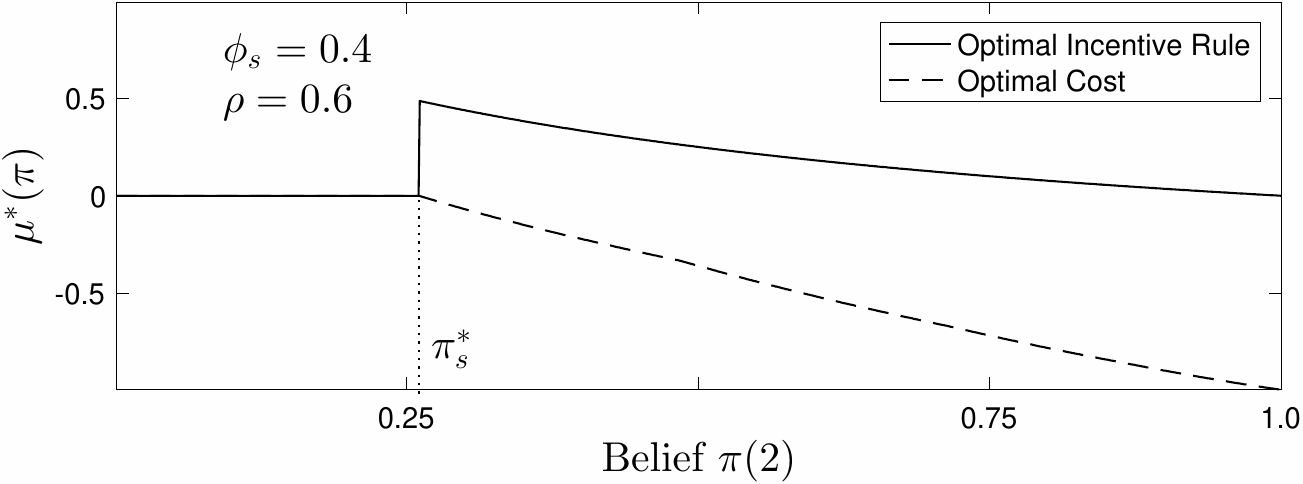}
        \caption{Optimal Policy for discount factor $\rho = 0.6$.}
        \label{subfig2:rho2}
    \end{subfigure}
    \caption{Optimal cost vs Discount factor. It is seen that a higher discount factor leads to smaller (expected) costs for higher states. This indicates that it is beneficial for the fusion center to attach more importance to future costs as as it should also take into account the benefit from sensors performing social learning. The parameters of the incentive function (\ref{eq:Delt}) are specified in Table~\ref{eq:Para} and the weight $\phi_s = 0.4$. Here $\rho$ denotes the discount factor in the objective (\ref{eq:RM}) and $\phi_s$ denotes the weight in the information fusion cost (\ref{eq:cst_es}). }
\label{fig:Tr_ds}
\end{figure*}
\subsection{Multi-threshold Incentive Policies} \label{subsec:MTP}

This subsection illustrates numerically the nature of the optimal incentive policies for formulations of the information cost more general than (\ref{eq:cst_es}), in particular we consider the {\em entropy} cost. We will see that the optimal incentive policy has a multi-threshold structure (as in Fig.\ref{subfig2:STMT}). 

 \textit{\underline{Expenditure \& Entropy Cost for Information Fusion}}:  Suppose the fusion center aims to minimize the expenditure to receive truthful accounts of the information gathered by the social sensors in addition to minimizing the entropy of the state estimate, i.e,
\begin{equation} \label{eq:cst_EE}
c(p) =  p + \psi_e(\pi) C_e(\pi) - \phi_e \Indc(a=y | \pi)
\end{equation}
where $\phi_e \in (0,1)$ denotes the scalar weight, $p$ denotes the expenditure, $\psi_e$ denotes the importance of the entropy cost, and $C_e(\pi) = -\sum_{i=1}^2 \pi(i) \text{log}_2 \pi(i) $~for $\pi(i) \in (0,1)$ and $C_e(\pi) {\overset{\Delta}=} 0$ for $\pi(i) = \{ 0,1\}.$ Fig.\ref{fig:EntrC} shows the optimal cost and optimal policy for the fusion center when it considers entropy of the state estimate in addition to the expenditure in the information fusion cost (\ref{eq:cst_es}). It can be seen that the optimal policy has a multi-threshold structure, and the optimal cost is discontinuous. A discontinuous cost implies a slight change in the initial conditions will lead to significantly different costs. Optimal policy being multi-threshold is unusual: it implies that if it is optimal to incentivize at a particular belief, it need not be optimal to do the same when the belief is larger. 
\begin{figure*}[t!]
    \begin{subfigure}[t]{0.48\textwidth}
       {\hspace{-0.2cm}} \includegraphics[width=6.5cm,height=2.5cm]{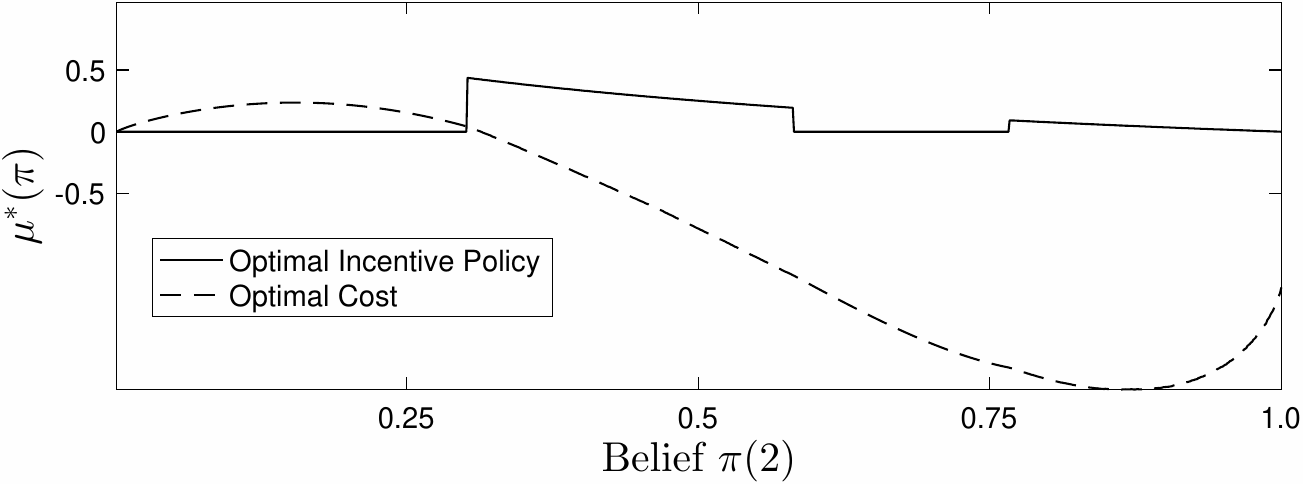}
        \caption{ The parameters are in Table~\ref{eq:Para} with $\phi_e = 0.25$, discount factor~$\rho = 0.8$, and~$\psi_e(\pi) = 0.1 - \pi^2(2)$. Here $\psi_e(\pi)$ captures the requirement of higher weight when the belief is smaller.}
        \label{subfig1:rho1}
    \end{subfigure}%
    ~~~
    \begin{subfigure}[t]{0.48\textwidth}
       \includegraphics[width=6.5cm,height=2.5cm]{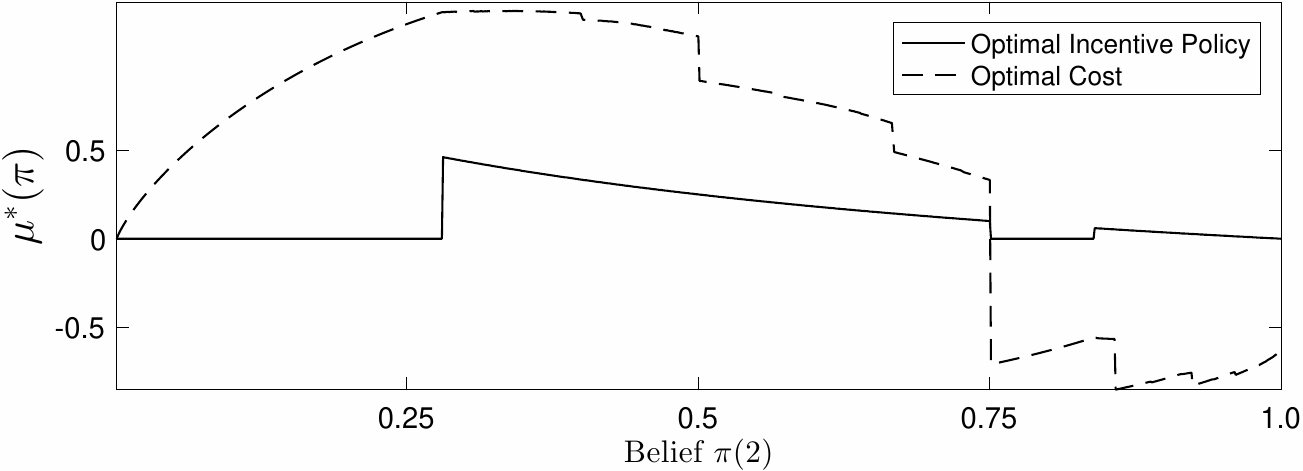}
        \caption{ Discontinuous optimal cost. The parameters are in Table~\ref{eq:Para} with~$\phi_e = 0.4$, discount factor~$\rho = 0.6$, and~$\psi_e(\pi) = 0.6 \times \Indc(\pi(2)<0.75) - 0.35  \times \Indc(\pi(2)>0.75)$. Here $\psi_e(\pi)$ captures the requirement of higher weight when the belief is smaller.}
        \label{subfig2:rho2}
    \end{subfigure}
    \caption{Multi-threshold incentive policy with the entropy cost. The regions in the belief space $\Pi(2)$ where it is optimal to not incentivize $\mu^*(\pi)=0$ is no more connected and convex. Having a connected region in the belief space where it is optimal not to incentivize has implications on the confidence of the fusion center in implementing the incentive policy: once its optimal to incentivize at a certain belief, it need not be optimal to continue incentivizing when the belief is larger, i.e, when it is more certain about the estimate of the state. The optimal cost is discontinuous in Fig.$7\text{b}$, and this implies that a slight change in the initial conditions will lead to a significantly different cost. }
\label{fig:EntrC}
\end{figure*} 

\subsection{Sensitivity of Optimal Incentive Policy} \label{subsec:SOMP}
The following numerical results along with Theorem~\ref{thm:AHI} provide a rationale for choosing the parameters: $\phi_s$ -- the weight in the information fusion cost (\ref{eq:cst_es}) and $\rho$ -- the discount factor in the fusion center's objective (\ref{eq:RM}).

(i)~\textit{\underline{Usefulness of Information vs Incentivizing}}: \\
We illustrate the trade-off between usefulness of information and incentivizing in the information fusion cost~(\ref{eq:cst_es}), and see how it affects the threshold $\pi^*_s$ in~(\ref{eq:OPESM}). Fig.\ref{fig:Tr_phi} shows the affect of increasing the weight $\phi_s$ when the remaining parameters are the same. It can be seen that $\pi^*_s$ is decreasing with $\phi_s$. 
From Theorem~\ref{thm:AHI}, higher $\phi_s$ implies that the additional cost for employing a sub-optimal policy is smaller; in other words, $\pi_s^*(2)$ is smaller.\\
(ii)~\textit{\underline{Optimal cost vs Discount factor}}: \\
We illustrate the relation between total cost incurred by the fusion center for different discount factors $\rho$ in the objective function (\ref{eq:RM}). The discount factor models the degree of impatience of the fusion center, as the cost incurred at time $k$ is $\rho^k c(p_k)$. A smaller discount factor indicates that the fusion center pays more attention to the current costs than future costs. It is seen from Fig.\ref{fig:Tr_ds} that a higher discount factor leads to smaller (expected) costs for higher states. This indicates that it is beneficial for the fusion center to attach more importance to future costs as it should also take into account the benefit from sensors performing social learning.
\begin{figure*} [t!]
\centering
 {\hspace{-0.2cm}}\includegraphics[width=9cm,height=3.5cm]{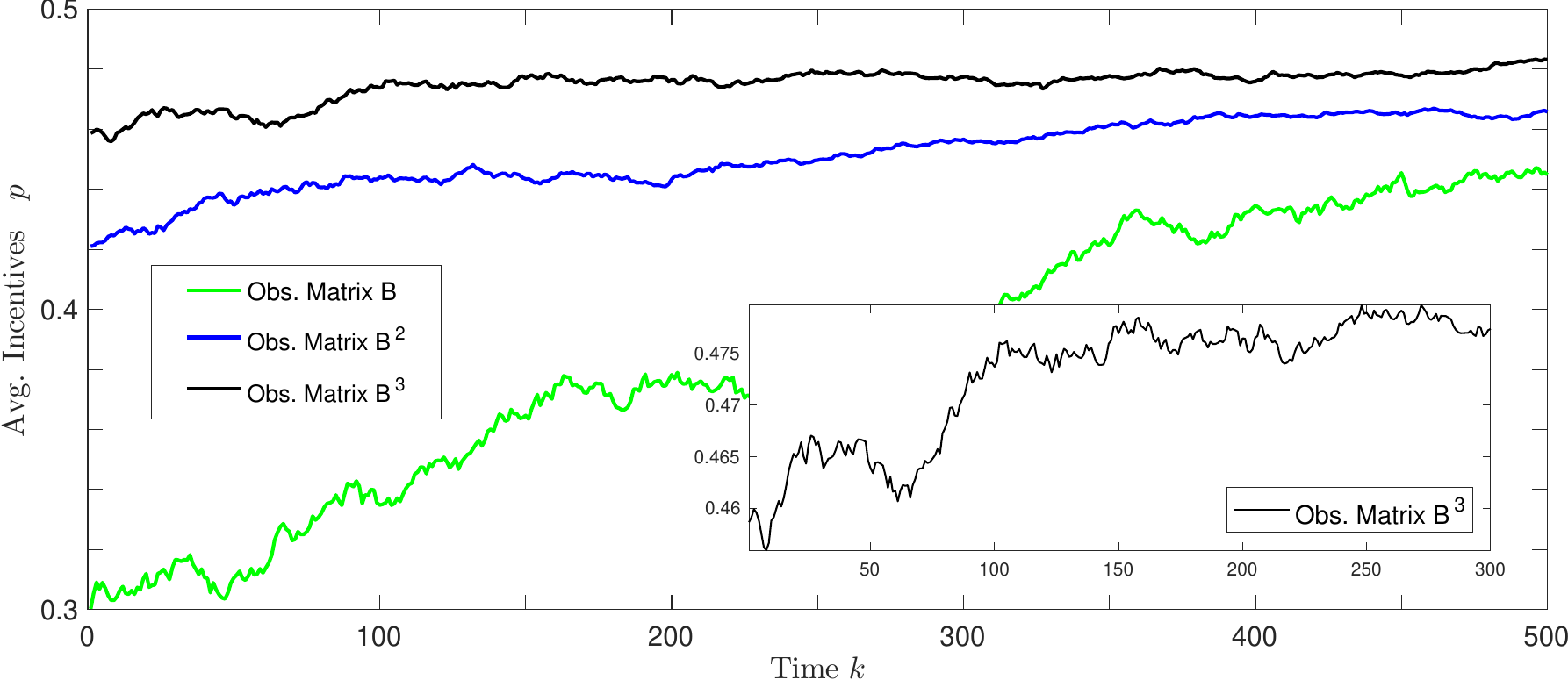}
 \caption{The figure shows the incentives averaged over independent sample paths for the fusion center over time for observation matrices $B$, $B^2$ and $B^3$. The observation matrices are ordered in the decreasing order of informativeness (see Footnote $8$). The parameters are specified in Tables I \& II. The weight $\phi_s=0.4$ in the information fusion cost (\ref{eq:cst_es}) and the discount factor $\rho = 0.6$. It can be seen that the range (or the slope) of the average incentives over the time horizon is highest for the case of observation matrix $B$. The average incentives display an increasing trend.  The zoomed in subfigure shows the increasing trend in case of observation matrix~$B^3$. It can be seen that average incentives offered in case of $B^3$ is higher than $B^2$ which in turn is higher than $B$.}
 \label{fig:avg_cst}
 \end{figure*}
\subsection{Sample Path of Optimal Incentives} \label{subsec:SPOI}
This subsection illustrates the sample path properties of the optimal incentive sequence over time (which was characterized in Theorem~\ref{thm:SMP} to be a sub-martingale). Fig.\ref{fig:avg_cst} shows the average incentives provided to the social sensors over time. The fusion center employs the optimal incentive policy~(\ref{eq:OPESM}) and fuses the information revealed by social sensors in a Bayesian way~(\ref{eq:SLF}). Each sample path has a duration of $N=500$, i.e, sequential information fusion from $500$ social sensors. The figure shows the average over $100$ independent such sample paths for three different observation likelihood matrices (\ref{eq:obs_m}). 
\begin{table} [t!]
	\begin{center}
		\begin{tabular}{ |c|c|c|c| } 
			\hline
			\multirow{2}{6em}{Obs. matrix~$B^2$} & $\alpha_1 = 0.3132$ & $\alpha_2 = 0.3032$ & $\beta_1 = 0.11$ \\ \cline{2-4}  
			& $ \beta_2 = 0.1$ & $\gacq_1 = 0.1$ & $\gacq_2 = 0.414$ \\
			\hline
			\multirow{2}{6em}{Obs. matrix~$B^3$} & $\alpha_1 = 0.3233$ & $\alpha_2 = 0.3133$ & $\beta_1 = 0.11$ \\ \cline{2-4}
			& $ \beta_2 = 0.1$ & $\gacq_1 = 0.1$ & $\gacq_2 = 0.414$ \\
			\hline
		\end{tabular}
	\end{center}
	\caption{The reward vector (\ref{eq:res_d_rwrd}) parameters for $B^2$ and $B^3$. For $\delta_1 = 0.3, ~\delta_2 = 0.95$, the following parameters were obtained as a solution of $\Delta(e_1)=1$ and $\Delta(e_2)=0$ for the reward vector (\ref{eq:res_d_rwrd}) parameters with observation matrix $B$=\usebox{\smlmat}.  }
	\label{tb:B2}
\end{table}
We consider the following observation likelihood matrices for illustrating the relation between the information gathering capabilities of the sensor (\ref{eq:obs_m}) and the average incentives provided by the fusion center: $B$, $B^2$, and $B^3$. We know that $B$ is more informative than $B^2$, which is in turn more informative than $B^3$, in the Blackwell sense \cite{Kri16} (see also  Footnote~$14$). 
\subsubsection*{\underline{Parameters}}\
The parameters of the incentive function (\ref{eq:Delt}) using the resolution dependent reward (\ref{eq:res_d_rwrd}) for $B^2$ and $B^3$ are specified in Table~\ref{tb:B2}. 
In Fig.\ref{fig:avg_cst}, it can be seen that the range (or the slope) of the average incentives over the time horizon is highest for the case of observation matrix $B$ (compared to $B^2$ and $B^3$). It can be seen from Fig.\ref{fig:avg_cst} that the average incentives display an increasing trend. 
 %
\subsection{Controlled Information Fusion in non-binary environments} \label{subsec:CFNB}
In this section, we briefly discuss the formulation for multiple states. Partial results on social learning with multiple states and $2$ actions appears in \cite{Kri12}. In the controlled fusion problem considered in this paper, the social sensors reveal the observation to the fusion center. This requires that the cardinality of $\mathcal{A}$ and $\mathcal{Y}$ be equal. Due to the complexity of analyzing the structural results for the optimal policy in case of multiple actions and states, we only describe the formulation and illustrate the incentive policy using a numerical simulation for a $\mathcal{X} = \mathcal{A} = \mathcal{Y} = \{1,2,3\}$. When $|\mathcal{X}|=3$, the public belief is in the belief space 
\begin{equation*}
\hspace{-1cm} \Pi(3) = {\overset{\Delta}{=}}\lbrace \pi \in \mathbb{R}^{2} : \sum_i \pi(i)= 1, 0 \leq \pi(i) \leq 1 ~\text{for} ~ i \in \{1,2,3 \}\}.
\end{equation*}
The number of regions in the space $\Pi(3)$ that need be considered for analyzing the structural results of the optimal incentive policy are $5$ (see (\ref{eq:REG_3}) below) as opposed to $3$ in (\ref{eq:REG}). \\
\underline{Model Assumptions}:
\begin{compactenum}
	\item[(A'1)] The observation distribution $B_{xy} = \mathbb{P}(y|x)$ is TP2 (totally positive of order 2), i.e, all second order minors of matrix $B$ are non-negative.
	\item[(A'2)] The reward vector $r_a$ is supermodular, i.e, $r_{a+1} - r_{a}$ is an increasing vector for $a = \{1,2\}$ and every $p \in [0,1]$.                                                                          
\end{compactenum}

The social sensors' decision $a(\pi,y) = {\text{arg~max}} ~r_a^{\prime}  \eta_{y}$ is increasing in $\pi$ and $y$ under (A'1) and (A'2); see \cite{Kri16}. This can be used to establish the single crossing condition, 
\begin{equation}
\hspace{-0.0cm}\{ \pi \in \Pi(3): (r_a - r_{a+1})^{\prime} \eta_{y}  \leq 0 \} \subseteq \{ \pi \in \Pi(3): (r_a - r_{a+1})^{\prime} \eta_{y+1}  \leq 0 \}.
\end{equation}
We now can define the following regions in the belief simplex $\Pi(3)$ (compare with (\ref{eq:REG})): 
\begin{align} \label{eq:REG_3}
\mathcal{P}_1^\cntrl &= \{ \pi \in \Pi(3): (r_1 - r_3)^{\prime} \eta_{y=1}  \cap (r_2 - r_3)^{\prime} \eta_{y=1}  \leq 0 \},  \nonumber\\ 
\mathcal{P}_2^\cntrl &= \{  \pi \in \Pi(3): (r_1 - r_3)^{\prime} \eta_{y=2}  \leq 0  \cap (r_2 - r_3)^{\prime} \eta_{y=2}  \leq 0 \nonumber\\ ~&\cap~  (r_1 - r_2)^{\prime} \eta_{y=1}  \leq 0\}, \nonumber \\ 
\mathcal{P}_3^\cntrl &= \{  \pi \in \Pi(3): (r_1 - r_3)^{\prime} \eta_{y=3}  \leq 0  \cap (r_2 - r_3)^{\prime} \eta_{y=3}  \leq 0 \nonumber\\ &~\cap (r_1 - r_2)^{\prime} \eta_{y=2}  \leq 0 \cap (r_2 - r_3)^{\prime} \eta_{y=2}  > 0 \cap~ (r_1 - r_2)^{\prime} \eta_{y=1}  > 0 \nonumber \\ &\cap \{(r_1 - r_3)^{\prime} \eta_{y=1}  > 0\}, \nonumber \\ 
\mathcal{P}_4^\cntrl &= \{  \pi \in \Pi(3): (r_1 - r_2)^{\prime} \eta_{y=3}  \leq 0  \cap (r_2 - r_3)^{\prime} \eta_{y=3}  > 0 \nonumber \\ ~&\cap~  (r_1 - r_2)^{\prime} \eta_{y=2}  > 0 \cap (r_1 - r_3)^{\prime} \eta_{y=2}  > 0\}, \nonumber \\ 
\mathcal{P}_5^\cntrl &= \{  \pi \in \Pi(3): (r_1 - r_2)^{\prime} \eta_{y=3}  > 0  \cap (r_1 - r_3)^{\prime} \eta_{y=3}  > 0 \}.
\end{align}

The value function for the fusion center is given by:
\begin{align} \label{eq:Val_Iter}
V(\pi) &= \min \{ c(p) + \rho \sum_a \sum_{j=1}^{5} V(T^{j}(\pi,a)) \sigma(\pi,a) \Indc(\pi \in \mathcal{P}_j^\cntrl)  \}, \nonumber \\
V(\pi) &= \underset{p \in [0,1]}{\min} \Big\{ p - \phi_s \Indc(\pi \in \mathcal{P}_3^\cntrl) + \nonumber \\ &\rho \sum_a \sum_{j=1}^{5} V(T^{j}(\pi,a)) \sigma(\pi,a) \Indc(\pi \in \mathcal{P}_j^\cntrl)  \Big\}.
\end{align}
Here $T^{j}(\pi,a) = \frac{R^{j}_a \pi}{\boldsymbol{1}^\prime R^{j}_a \pi}$, with $R^{j} = BM^{j}$ for $j = 1,2,\cdots,5$.\\
Fig.\ref{fig:Mult_st} shows the optimal incentive policy for a $3$ state, observation, and action model. Lemma~$\ref{lem:IncPB}$ in the paper can be used to find the matrices $M^{j}$ for $j = 1,2,\cdots,5$. 
The observation distribution for the controlled fusion problem for $3$ states and actions is chosen as:
\[B = \begin{bmatrix}
0.7479  &  0.1986  &  0.0536 \\
0.6023  &  0.2543   & 0.1434 \\
0.2785  &   0.2459   &  0.4756
\end{bmatrix}.\]
\begin{figure}[t!]
	\centering
	{\hspace{0cm}}\includegraphics[scale=0.12]{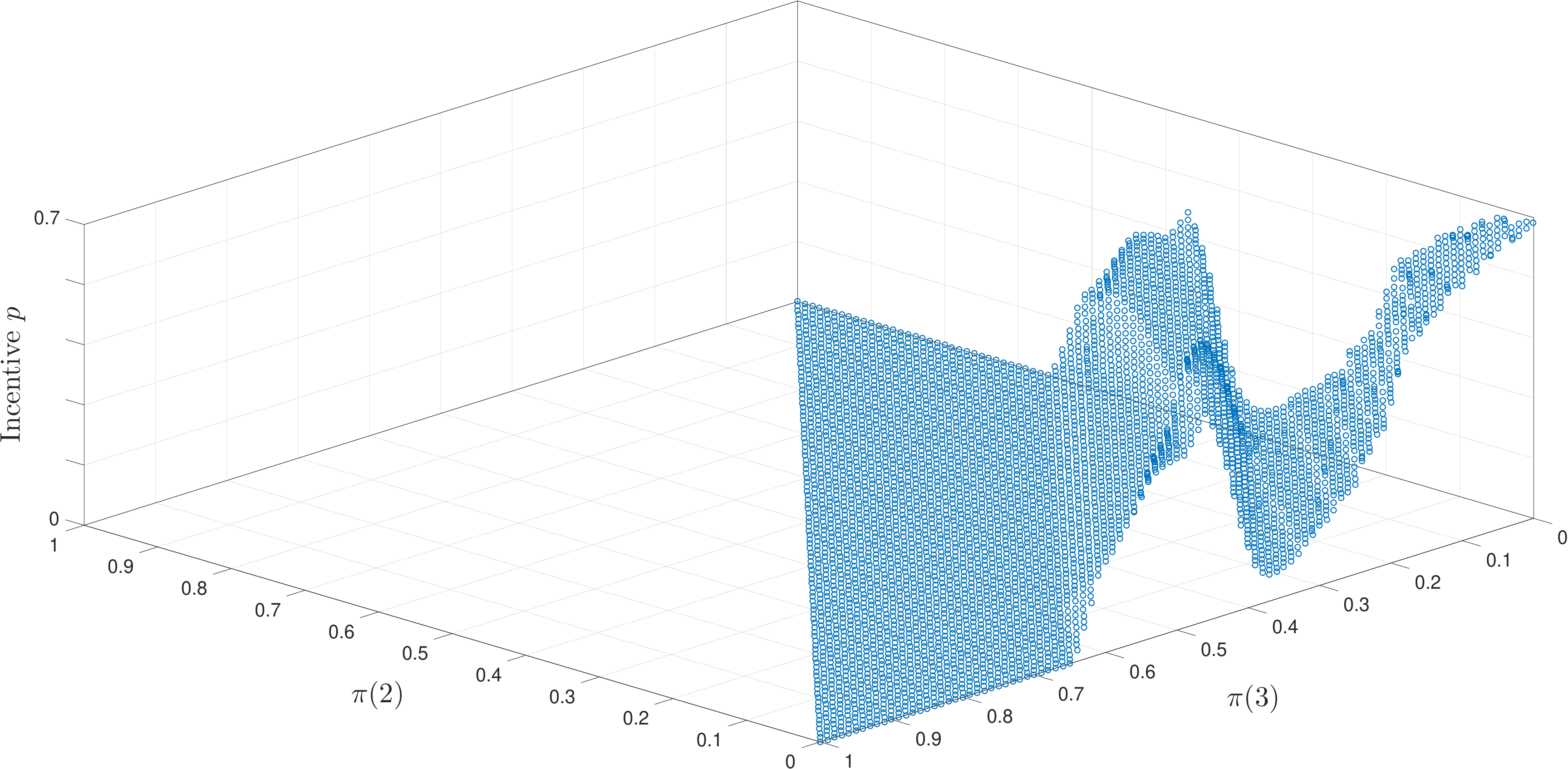}
	\caption{Optimal Incentive Policy for social learning weight $\phi_s = 0.6$, $\rho = 0.8$, $\beta_1 = 0.6771, \beta_2 = 0.5465, \beta_3 = 0.7113$, $\delta_1 = 0.3, \delta_2 = 0.4, \delta_3 = 0.5$, $\gamma_1 = 0.5, \gamma_2 =  0.3, \gamma_3 = 0.2$, and $\alpha_a = 0$ for all $a \in \{1,2,3\}$. The belief space $\Pi(3)$ was discretized into a grid of $5151$ belief points using Fruedenthal triangulation \cite{Kri16}. The incentive $p \in [0,1]$ was discretized into~$50$ values.}
\label{fig:Mult_st}
\end{figure}
The value iteration algorithm based on (\ref{eq:Val_Iter}) was run for a horizon~$N=100$.

\section{Conclusion and Future Work}
Unlike data fusion involving physical sensors for tracking targets, this paper is motivated by information fusion with social sensors, which provide reviews on social media review platforms such as Amazon, Yelp, and Airbnb. Our main objective is to control the information fusion by dynamically providing incentives to the social sensors. We presented five main results. Theorem~\ref{thm:OPRM} showed that under reasonable conditions on the model parameters, the optimal incentive policy has a threshold structure. The optimal policy is determined in closed form, and is such that it switches once between two exactly specified incentive policies. Theorem~\ref{thm:SMP} characterized the sample path property of the optimal incentive sequence that results from fusion center employing the optimal threshold policy. It was shown that the optimal incentive sequence is a sub-martingale. Theorem~\ref{lem:lmts} showed how the fusion center can employ a sub-optimal policy and thereby facilitate social learning indefinitely, to learn the true state asymptotically. In other words, it was shown how controlled information fusion with social sensors can be consistent. Theorem~\ref{thm:AHI} provided uniform bounds on the average additional cost incurred, by employing a sub-optimal policy, for consistency. Theorem~\ref{thm:AHF} provided uniform bounds on the budget saved by employing a policy that estimates the state with a degree of confidence, instead of the optimal policy. Finally, Theorem~\ref{thm:SETr} established that the optimal policy for estimating a random variable is near optimal for tracking a changing state, when out-of-state transition probabilities are small. 

While the formulation of the controlled information fusion problem applies to arbitrary finite state, observation and action spaces, our structural analysis of the optimal incentive policies are currently applicable only to the $2$ state case. We briefly discussed the formulation for the case of $3$ states, observations and actions, and highlighted the difficulty in deriving structural results in non-binary environments.

%
\appendix
\section{Definitions and Preliminaries:}
\begin{definition}{\textit{First-Order Stochastic Dominance (FSD)} ($\geq_{s}$):} Let $\pi_{1},\pi_{2}\in \Pi(2)$ be any two belief state vectors. Then $\pi_{1}\geq_{s}\pi_{2}$ if 
\begin{equation} \label{def:fodm}
{\overset{2}{\underset{i=j}{\sum}}} \pi_{1}(i) \geq {\overset{2}{\underset{i=j}{\sum}}} \pi_{2}(i) ~ \text{for} ~ j \in \{1,2\}.
\end{equation}  
Equivalently, $\pi_{2} \geq_{s} \pi_{1}$ iff for all $v\in \mathcal{V}$, $v'\pi_{2}\leq v'\pi_{1}$, where $\mathcal{V}$ denotes the space of $2$-dimensional vectors $v$, with non-increasing components, i.e, $v_{1} \geq v_{2} \geq \hdots v_{X}$.
\end{definition}
\begin{definition}{(Martingale~\cite{Dur10}):} 
Let $\mathcal{F}_k$ denote the sigma algebra (as in (\ref{eq:SgF})). A sequence $\{X_k\}$ such that $\mathbb{E}[| X_k|] < \infty$ is a martingale (with respect to $\mathcal{F}_k$) if 
$$ \mathbb{E}[X_{k+1} | \mathcal{F}_k] = X_k, ~\text{for all}~k.$$
\end{definition}
If $ \mathbb{E}[X_{k+1} | \mathcal{F}_k] \geq X_k, ~\text{for all}~k.$, the sequence $\{X_k\}$ is a \textit{sub-martingale}.
\begin{definition} (\cite{Dur10})
A sequence $H_k$ is said to be a predictable sequence if $H_k \in \mathcal{F}_{k-1}$.
\end{definition}
In words, $H_k$ may be predicted with certainty using the information available at time $k-1$. 
\begin{lemma}[\cite{Kri16}]\label{lem:FDs}
Under (A1), we have $\sigma(\pi_1,a) \geq_s \sigma(\pi_2,a)$, where 
$\sigma(\pi,a) = \begin{bmatrix}  
\textbf{1}'B_{y=1}^{\pi} \pi \\  
\textbf{1}'B_{y=2}^{\pi} \pi
\end{bmatrix}$. 
\end{lemma}

%
\begin{lemma}[\cite{Kri12}] \label{lem:IncPB}
The sensor decision likelihood probability matrix $R^{\pi}$ in the social learning filter (\ref{eq:SLF}) is computed as
\begin{align} \label{eq:BReg}
R^{\pi} &= B M^{\pi} ~\text{where}~
M_{y,a}^{\pi} = \mathbb{P}(a | y,\pi) = ~\Indc(r_a^{\prime} B_y  \pi > r_{\bar{a}}^{\prime} B_y  \pi ), \nonumber \\ ~&\text{with}~ \bar{a} = \mathcal{A}/a.  
 \end{align}
\end{lemma}

\begin{theorem}[\cite{Kri12}] \label{thm:VFRM}
Let (A1) and (A2) hold. The belief space $\Pi(2)$ can be partitioned into at most $3$ non-empty regions $\mathcal{P}_1, \mathcal{P}_2, \mathcal{P}_3$. On each of these regions, the sensor decision likelihood matrix $R^{\pi}$ in (\ref{eq:BReg}) is a constant with respect to the belief state~$\pi$.
\end{theorem}
%
%

%
\begin{theorem}[ \cite{Dur10}] \label{thm:PSM}
Let $W_k$ be a sub-martingale. If $H_k \geq 0$ is predictable and each $H_k$ is bounded, then $(H.W)_k$ is a sub-martingale. 
\end{theorem}
Theorem~\ref{thm:PSM} corresponds to Theorem~$5.2.5$ in \cite{Dur10}.
%
%
%
\section{Proofs}
{\bf{\underline{Proof of Theorem~\ref{thm:OPRM}}}}: \\
We first show that due to the structure of the social learning filter in (\ref{eq:SLF}), the choice of incentives reduces from a continuum $[0,1]$ to a finite number at every belief. 
Next, we show that the incentive function $\Delta(\eta_y)$ is decreasing in $\pi$ for any $y$. 
\begin{theorem} \label{thm:RM}
Let $\Delta(\eta_{y=1})$ and $\Delta(\eta_{y=2})$ be two possible incentives at belief $\pi$. Under (A1) and (A2), the $Q$ function in (\ref{eq:RMVP}) can be simplified as:\\
\begin{equation} \label{eq:ESM}
\hspace{-0cm} Q(\pi,\cntrl) = \left\{ \begin{array}{lll}
\cntrl + \rho V(\pi) & \mbox{if $\cntrl \in [0,\Delta(\eta_{y=2}))$}; \\
       \cntrl  -\phi_s   + \rho \E V(\pi) & \mbox{if $\cntrl \in [\Delta(\eta_{y=2}), \Delta(\eta_{y=1}))$}; \\
      \cntrl +\rho V(\pi)  & \mbox{if $\cntrl \in [\Delta(\eta_{y=1}), 1]$}.\end{array} \right. 
 \end{equation}    
and $V(\pi) = \min Q(\pi,\cntrl)$. 
%
Here,
$
\E V(\pi) = \textbf{1}'B_{y=1}^{\pi} \pi \times V(\eta_{y=1}) + \textbf{1}'B_{y=2}^{\pi} \pi \times V(\eta_{y=2}).
$
\end{theorem}
\textit{\underline{Proof of Theorem~\ref{thm:RM}}}: \\

From Lemma~\ref{lem:lmts} and Theorem~\ref{thm:VFRM}, we have
\begin{equation} \label{eq:DeM}
\hspace{-0cm} R^{\pi} = \left\{ \begin{array}{lll}
        \begin{bmatrix}
1 & 0\\
1 & 0
\end{bmatrix} & \mbox{if $p  \in [0,\Delta(\eta_{y=2}))$};\\
       \begin{bmatrix}
B_{11} & B_{12}\\
B_{21} & B_{22}
\end{bmatrix} & \mbox{if $p  \in [\Delta(\eta_{y=2}), \Delta(\eta_{y=1}))$}; \\
       \begin{bmatrix}
0 & 1\\
0 & 1
\end{bmatrix} & \mbox{if $p  \in [\Delta(\eta_{y=1}), 1]$}.\end{array} \right. 
 \end{equation}   

From (\ref{eq:DeM}), it is clear that the sensors' decision
\begin{equation} 
\hspace{-0cm} a = \left\{ \begin{array}{lll}
         1 & \mbox{if $p  \in [0,\Delta(\eta_{y=2}))$};\\
        y & \mbox{if $p  \in [\Delta(\eta_{y=2}), \Delta(\eta_{y=1}))$}; \\
       2 & \mbox{if $p  \in [\Delta(\eta_{y=1}), 1]$}.\end{array} \right. 
 \end{equation}  

Therefore, 
\begin{equation} \label{eq:CCst}
\hspace{-0cm} \sum_{a \in \mathcal{A}} V(T^{\pi} (\pi,a)) \sigma(\pi,a) = \left\{ \begin{array}{lll}
V(\pi) & \mbox{if $p \in [0,\Delta(\eta_{y=2}))$};\\
       \mathbb{E} V(\pi) & \mbox{if $p  \in [\Delta(\eta_{y=2}), \Delta(\eta_{y=1}))$}; \\
      V(\pi) & \mbox{if $p \in [\Delta(\eta_{y=1}), 1]$}.\end{array} \right . 
 \end{equation}  
where $\E V(\pi) = \textbf{1}'B_{y=1}^{\pi} \pi \times V(\eta_{y=1}) + \textbf{1}'B_{y=2}^{\pi}  \pi \times V(\eta_{y=2})$.  \qed 

Theorem~\ref{thm:RM} represents the Q function (\ref{eq:RMVP}) over the range $[0,1]$ into \textit{three} regions. The following corollary highlights why such a partition is useful. 

\begin{corollary} \label{cor:FnP}
At every public belief $\pi \in \Pi(2)$, it is sufficient to choose one of the three incentives $\{ 0, \Delta(\eta_{y=2}), \Delta(\eta_{y=1})\}$.
\end{corollary} 

\textit{Proof}. From Theorem~\ref{thm:RM}, the instantaneous reward is a linear function in $\cntrl$ and 
\begin{align*}
\hspace{-1cm} \argmin_{\cntrl \in [0, \Delta(\eta_{y=2}))} Q(\pi,\cntrl) &= 0,\argmin_{\cntrl \in [\Delta(\eta_{y=2}), \Delta(\eta_{y=1}))} Q(\pi,\cntrl) = \Delta(\eta_{y=2}),\nonumber \\ \argmin_{\cntrl \in  [\Delta(\eta_{y=1}), 1]} &Q(\pi,\cntrl) = \Delta(\eta_{y=1}).
\end{align*} 
These hold as for any value of $\cntrl$ in each of the three regions, the corresponding continuation payoff is the same from Theorem~\ref{thm:RM}. \qed \\
\begin{lemma} \label{lem:D_dec}
The incentive function $\Delta(\eta_y)$ is decreasing in $\pi$ for every $y$.
\end{lemma}
\textit{Proof}. The incentive function is given as (\ref{eq:Delt}), where $l_1,l_2,l_3 > 0$.
With $\pi = [1 - \pi(2),~ \pi(2)]^\prime$, differentiating w.r.t $\pi(2)$, 
\begin{equation*}
\frac{d (\Delta(\eta_{y}))}{d \pi(2)} = -(l_1+l_2) B_{1y} B_{2y} < 0 \qed
\end{equation*}

\textit{\underline{Proof of Theorem~\ref{thm:OPRM}}}: \\
 From Corollary~\ref{cor:FnP} the value function (\ref{eq:RMVP}) is:
\begin{align} \label{eq:ESM_V}
\hspace{-1.1cm} V(\pi) &= \min\{  \rho V(\pi), \Delta(\eta_{y=2}) - \phi_s + \rho \mathbb{E} V(\pi), \Delta(\eta_{y=1})  + \rho V(\pi) \}. \nonumber\\
\hspace{-1.1cm}  \Rightarrow  V(\pi) &= \min\{  0, \Delta(\eta_{y=2}) - \phi_s + \rho \mathbb{E} V(\pi)\},    
\end{align}
as $\Delta(\eta_{y=1}) \geq 0$. \\
By using the value iteration algorithm \cite{Kri16} on (\ref{eq:ESM_V}), we have
\begin{align}
V_{n+1}(\pi) &= \min\{  0, \Delta(\eta_{y=2}) - \phi_s + \rho \mathbb{E} V_{n}(\pi)\}
\end{align}
with $V_0(\pi) = 0~\forall~\pi$. \\  
From Lemma~\ref{lem:D_dec}, the incentive function is decreasing. From the definition of First-Order Stochastic Dominance~(\ref{def:fodm}), and Lemma \ref{lem:FDs}, we have 
$\mathbb{E} V_{n}(\pi)$ is decreasing in $\pi$. Therefore, $V_{n+1}(\pi)$ and hence $V(\pi)$ is decreasing in $\pi$. \\
Let $V(0)$ and $V(1)$ denote the values for $\pi = \left\lbrace \begin{bmatrix}  1 \\ 0 \end{bmatrix}, \begin{bmatrix}  0 \\ 1 \end{bmatrix} \right\rbrace$. It is seen by substitution that $\mathbb{E} V(0) = V(0)$ and $\mathbb{E} V(1) = V(1)$. 
%
By definition, we know that $\Delta(\eta_y) \in [0,1]$. Using{\footnote{Note that after normalization $\Delta(e_1) = 1$ and $\Delta(e_2)=0$.}} Lemma~\ref{lem:D_dec}, let $\Delta(e_1) > \phi_s$ and $\Delta(e_2)< \phi_s$. The value function for the fusion center is given by (\ref{eq:ESM_V}). We have the following:
\begin{compactenum}
\item For $V(\pi) =  \Delta(\eta_{y=2})  - \phi_s + \rho \mathbb{E} V(\pi)$,  $V(0) = \frac{\Delta(e_1) - \phi_s}{(1-\rho)}>0$, and $V(1) = \frac{\Delta(e_2)- \phi_s}{(1-\rho)} < 0$. 
\item For $V(\pi) =  0$, $V(0) = V(1)  = 0$. 
\end{compactenum}
The value function $V(\pi)$ in (\ref{eq:ESM_V}) is decreasing with a positive value at $e_1$ and a negative value $e_2$, so must be zero at some point(s). Let $\Sigma =\{ \pi(2) | 0 = \Delta(\eta_{y=2})  - \phi_s + \rho \mathbb{E} V(\pi) \}$. Since the value function $V(\pi)$ is monotone in $\pi$, the set $\Sigma$ is convex. \\
Choosing $\pi_s^*(2) = \{ \hat{\pi}(2) | \hat{\pi}(2) > \pi(2)~\forall~\pi(2) \in \Sigma\}$, the result follows. \qed 

\vspace{0.1cm}
{\bf{\underline{Proof of Theorem~\ref{thm:SMP}}}}: \\
We will first establish the property of the incentive function $\Delta(\eta_y)$ as the optimal policy depends on it. 
\begin{lemma} \label{lem:PL}
Under (A1), $\Delta(\eta_{y=1})$ is concave in $\pi$, and $\Delta(\eta_{y=2})$ is convex in $\pi$.
\end{lemma}
{\textit{\underline{Proof of Lemma~\ref{lem:PL}}}}: \\
The incentive function $\Delta(\eta_{y=2})$ is given in (\ref{eq:Delt}). A differentiable function $f: [0,1] \rightarrow [0,1]$ is convex if
\begin{equation} \label{eq:Cnvx}
f(w_1) \geq f(w_2) +  f^{\prime}(w_2) (w_1 - w_2), \text{for all}~w_1,w_2 \in [0,1].
\end{equation}
A function $f$ is concave if $-f$ is convex.\\
%
From (\ref{eq:Cnvx}) with $w_1 = \pi_1(2)$ and $w_2 = \pi_2(2)$, and using Lemma~\ref{lem:FDs}, it is verified that the function $\Delta(\eta_{y=2})$ is convex in $\pi$. Similarly, it can be shown that $\Delta(\eta_{y=1})$ is concave in~$\pi$. \qed \\
{\textit{\underline{Proof of Theorem~\ref{thm:SMP}}}}: \\
Consider the sub-optimal policy $\hat{\mu}(\pi)$  given as
\begin{equation*} 
\hat{\mu}(\pi) = \left\{ \begin{array}{lll}
          \Delta(\eta_{y=2}) - \epsilon & \mbox{if $\pi(2) \in [0,\pi_*(2))$};\\
       \Delta(\eta_{y=2}) & \mbox{if $\pi(2) \in [\pi_*(2), 1]$}.\end{array} \right. 
        \end{equation*}   
   Here $\epsilon > 0$ and $\pi_*(2) \in [0,1]$. Let $W_k = \hat{\mu}(\pi_{k-1})$. \\
From Lemma~\ref{lem:PL}, $\Delta(\eta_{y=2})$ is convex in $\pi$. Let $u^{S}(\pi_{k+1}) = \Delta(\eta_{y_k=2})$ denote the price at time $k+1$. So $u^{S}(\pi)$ is convex in~$\pi$. \\
We know that the public belief $\pi_k$ is a martingale (\cite{Cha04}), i.e, $\E[\pi_{k+1}| \mathcal{F}_k] = \pi_k$. For $\epsilon \rightarrow 0$,  
\begin{equation*}
\hspace{-0cm} \E[W_{k+1}| \mathcal{F}_k] = \E[u^{S}(\pi_{k+1})| \mathcal{F}_k] \geq u^{S}(\E[\pi_{k+1}| \mathcal{F}_k])  \geq u^{S}(\pi_k) \geq W_k
\end{equation*}
by Jensen's inequality and martingale property of the public belief. Therefore $W_k (=\hat{\mu}(\pi_{k-1}))$ is a sub-martingale.\\
Consider a function $\bar{\mu}(\pi)$ given by
\begin{equation*} 
\bar{\mu}(\pi) = \left\{ \begin{array}{lll}
         0 & \mbox{if $\pi(2) \in [0,\pi^*(2))$};\\
      1 & \mbox{if $\pi(2) \in [\pi^*(2), 1]$}.\end{array} \right. 
        \end{equation*}    
 Let $H_k = \bar{\mu}(\pi_{k-1})$.  From Theorem~\ref{thm:PSM}, $(H.W)_k$ is a sub-martingale. But $(H.W)_k = p_k$. 
Therefore, the optimal incentive sequence $\cntrl_k = \mu^*(\pi_{k-1})$ is a sub-martingale, $\E[\cntrl_{k+1} | \mathcal{F}_k] \geq \cntrl_k$,  i.e, it increases on average over time. \qed

{\bf{\underline{Proof of Theorem~\ref{lem:lmts}}}}: \\
We'll prove that $\pi \in \mathcal{P}_2^{p}$ iff $p  \in [\Delta(\eta_{y=2}), \Delta(\eta_{y=1}))$. Other cases are proved similarly. \\
We can write 
\begin{align}
\hspace{0cm} r_1 &= [(\delta_1 p - \gacq_1)  ~~ (\delta_1 p - \alpha_1 - \gacq_1)], \\ 
r_2 &= [(\delta_2 p - \alpha_2 - \gacq_2) ~~ (\delta_2 p - \gacq_2)]. 
\end{align}
By definition, 
\begin{equation*}
\hspace{-0cm} \mathcal{P}_2^\cntrl = \{  \pi \in \Pi(2): (r_1 - r_2)^{\prime} \eta_{y=1} > 0 ~\cap~(r_1 - r_2)^{\prime} \eta_{y=2} \leq 0 \}.
\end{equation*}
We have,
\begin{align*}
\hspace{-0cm} (r_1 - r_2)^{\prime}  \eta_{y=1} > 0  &\Leftrightarrow  p <  \frac{1}{\delta_2-\delta_1}  &\left\{ [\alpha_2 ~ -\alpha_1] \eta_{y=1} 
 +(\gacq_2 - \gacq_1)  \right \} \\ &= \Delta(\eta_{y=1}) . \\
\hspace{-0cm} (r_1 - r_2)^{\prime}  \eta_{y=2} \leq 0  &\Leftrightarrow p \geq  \frac{1}{\delta_2-\delta_1} & \left\{ [\alpha_2 ~ -\alpha_1] \eta_{y=2} +(\gacq_2 - \gacq_1)  \right \} \\ &= \Delta(\eta_{y=2}).  
\end{align*}

{\bf{\underline{Proof of Theorem~\ref{thm:AHI}}}}: \\
Define the following region in the belief space $\Pi(2)$:
\begin{equation} \label{eq:H_eq}
\mathcal{H} = \{ \pi | \pi(2) \leq \pi^*(2) \}. 
\end{equation}
 Here $\mathcal{H}$ denotes the region where the optimal policy in (\ref{eq:OPESM}) is such that $\mu^*(\pi) = 0$.
For any sub-optimal policy $\mu_c$ and the corresponding cost $W_{\mu_c}(\pi)$, it is clear that $W_{\mu_{c}}(\pi) - J_{\mu^*}(\pi) \geq 0 ~ \forall~\pi$. Also, $W_{\mu_{c}}(e_2) = J_{\mu^*}(e_2)$.  Let $\Indc$ denote the indicator function. We have
\begin{align} \label{eq:sup_WV}
{\hspace{-0.5cm}}W_{\mu_{c}}(\pi) - J_{\mu^*}(\pi) &= \Indc(\pi \in \mathcal{H}) \{ W_{\mu_{c}}(\pi) - J_{\mu^*}(\pi)\} \nonumber \\ &+ \Indc(\pi \notin \mathcal{H}) \{ W_{\mu_{c}}(\pi) - J_{\mu^*}(\pi) \} \nonumber \\
{\hspace{-0.5cm}} \Rightarrow \sup_\pi |W_{\mu_{c}}(\pi) - J_{\mu^*}(\pi) | &\leq \Big\{ \sup_\pi  \Indc(\pi \in \mathcal{H}) \{ W_{\mu_{c}}(\pi) - J_{\mu^*}(\pi)\} \Big\} \nonumber \\ &+ \Big\{ \sup_\pi  \Indc(\pi \notin \mathcal{H}) \{ W_{\mu_{c}}(\pi) - J_{\mu^*}(\pi)\} \Big\}.
\end{align}
where $\mathcal{H}$ is defined in (\ref{eq:H_eq}). From Theorem~\ref{thm:OPRM}, we know that $J_{\mu^*}(\pi)=V(\pi)$ is monotone (non-increasing) in $\pi$. Similar arguments can be used to establish that $W_{\mu_{c}}(\pi)$ is monotone (non-increasing) in $\pi$. Therefore, we have{\footnote{Equation (\ref{eq:Req_sup}) follows from $$ \mathbb{E}_{\mu_c} \Big\{ \sum_{k=0}^{\infty} \rho^k \{  c_{\mu_c}(p_k) \}  \Big\} >  \mathbb{E}_{\mu_c} \Big\{ \sum_{k=0}^{\infty} \rho^k \{ \Indc(\pi_k \in \mathcal{H})  c_{\mu_c}(p_k) \} \Big\}. $$}} for (\ref{eq:sup_WV})
\begin{align} \label{eq:Req_sup}
\sup_\pi |W_{\mu_{c}}(\pi) - J_{\mu^*}(\pi) | &\leq 2\Big\{ \sup_\pi \Indc(\pi \in \mathcal{H})  W_{\mu_{c}}(\pi)  \Big\} 
\end{align}
as $J_{\mu^*}(\pi) = 0~\forall~\pi \in \mathcal{H}$ from (\ref{eq:ESM_V}) and Theorem~\ref{thm:OPRM}. 
The set $\mathcal{H}$ defined in (\ref{eq:H_eq}) is compact by definition. For the discount factor $\rho \in [0,1)$ and bounded instantaneous costs, the cumulative discounted cost is bounded \cite{Kri16}. Therefore in (\ref{eq:Req_sup}), $$\sup_\pi \{ \Indc(\pi \in \mathcal{H})  W_{\mu_{c}}(\pi) \}  = \max_\pi \{ \Indc(\pi \in \mathcal{H}) W_{\mu_{c}}(\pi) \}$$
and~$\tilde{\pi} = \argmax_\pi \{  \Indc(\pi \in \mathcal{H}) W_{\mu_{c}}(\pi) \}$. We have for $\pi_0 = \tilde{\pi}$,
\begin{align*}
\hspace{-0cm}\max_\pi \{ \Indc(\pi \in \mathcal{H}) W_{\mu_{c}}(\pi) \}  & =  \mathbb{E}_{\mu_c} \Big\{ \sum_{k=0}^{\infty} \rho^k \{  c_{\mu_c}(p_k) \} \Big | \pi_0 = \tilde{\pi} \Big\} \\ &\leq \mathbb{E}_{\mu_c} \Big\{ \sum_{k=0}^{\infty} \rho^k  \max_{\Delta(\eta_{y=2}): \pi \in \mathcal{H}} c_{\mu_c}(p_k)  \Big\} \\ &= (1 - \phi_s) \mathbb{E} \Big\{ \sum_{k=0}^{\infty} \rho^k   \Big\} \\
\hspace{-1cm} & = \frac{(1 - \phi_s)}{1 - \rho}.  \qed
\end{align*}
\section{Discussion of Reward Functions} \label{App:Dis}
\subsection{Social Sensor's Reward Function}
The nature of the results, specifically, the structural results (Theorem~$1$ and Theorem~$3$ in Sec.\ref{sec:SOPP}); characterization of optimal incentive sequence (Theorem~$2$ in Sec.\ref{sec:SOPP}); and the uniform bounds (Theorem~$5$ and Theorem~$6$ in Sec.\ref{sec:CIF}); is unaffected by the choice of the form of reward functions below. 
\begin{itemize}
	\item[a.)] (Resolution dependent reward): This form of reward function can be used to explicitly capture the effect or the influence of the observation distribution (resolution) matrix $B$ of the social sensors on the actions. Let $r(x,y,a)$ denote the reward accrued if the sensor takes action $a$ when the underlying state is $x$ and the observation is $y$. The reward function is given as:
	\begin{align} \label{eq:res_d_rwrd}
	r(x,a) &= \sum_y r(x,y,a) B_{xy}, \nonumber \\ 
	\text{where}~r(x,y,a) &= \delta_a p - \alpha_a \Indc(a \neq x) - \beta_a \Indc(a \neq y) - \gacq_a.
	\end{align}
	
	Here $\delta_a \in [0,1]$,  $\alpha_a, \beta_a, \gacq_a \in \mathbb{R}$ are the given parameters of the model and $\Indc$ denotes the indicator function. For an action $a \in \mathcal{A}$ of the social sensor, $\delta_a p$ the effective incentive received (see discussion below) by the social sensor; $\gamma_a$  denotes the cost of taking the action; $\alpha_a$ and $\beta_a$ denote the mis-representation or distortion weights.
	\item[b.)] (Resolution independent reward): This form of the reward function is not explicitly dependent on the resolution of the social sensors, i.e,
	\begin{equation}
	r(x,a) = \delta_a p - \alpha_a \mathcal{I}(a \neq x)  - \gamma_a.
	\end{equation}
	\item[c.)] (Realization dependent reward): This form of reward function explicitly depends on the private observation or realization $y_k$ for the social sensor $k$, i.e,
	\begin{equation} \label{eq:rwd_rel}
	r(x,y_k,a) = \delta_a p - \alpha_a \Indc(a \neq x) - \beta_a \Indc(a \neq y_k) - \gacq_a.
	\end{equation} 
	\item[d.)]  (General state-action reward): This form of reward function models a general state-action reward function, i.e,
	\begin{equation} \label{eq:gen_Rwd}
	r(x,a) = \delta_a p + \Gamma^y_{xa}
	\end{equation}
	The parameter $\Gamma^y_{xa}$ is any function of the resolution, realization, state, and action. 
\end{itemize}
{\textit{Motivation}:} The social sensor $k$ receives a noisy observation $y_k$ of the state $x$. The term $\beta_a I(a \neq y_k)$ models the distortion cost \cite{OLMRR17} induced by the sensor's realization in equation (\ref{eq:rwd_rel}). 
For social sensor $k$, $I(a_k \neq y_k)$ is the binary distance function \cite{OLMRR17} of the distortion or mis-representation of the received information $y_k$ as $a_k$. In case of (\ref{eq:res_d_rwrd}), the term $\beta_a I(a \neq y)$ captures the inherent distortion that can result from the sensor's observation matrix $B$.\\
The information fusion center offers a single incentive $p_k \in [0,1]$ to the social sensor $k$ by using the information from the actions of the previous social sensors contained in the public belief $\pi_{k-1}$ (see (\ref{eq:Pol_tim})). The weight $\delta_a$ helps to model asymmetric incentives for the different actions of the social sensor, and determines the effective incentive received by the social sensor for choosing different actions. The asymmetry is required to derive a feedback (public belief dependent) policy for the information fusion center to choose the future price. Symmetry ($\delta_{a=2} =  \delta_{a=1}$) results in open loop or static prices (as the dependency cancels out) for the information fusion center. Since we are interested in dynamically changing the incentives to incorporate learning from the previous social sensors, we choose $\delta_{a=2} \neq  \delta_{a=1}$. 

\subsection{Information Fusion Cost}
The cost function for the fusion center is motivated by the revenue maximization problem with social learning literature \cite{BOOV06,Ott96,Cha04,BOOV08}: 
\begin{equation} \label{eq:Mon_Obj}
\sum_{k=0}^{\infty} \rho^k (p_k - c) \mathcal{I}(a_k=\text{buy}).
\end{equation}
Here (\ref{eq:Mon_Obj}) is the objective function of a monopoly that dynamically charges a price $p_k$ for a product that costs $c$ to manufacture, to a social sensor $k$ that learns about the underlying value (state) of the product from the decisions of other social sensors. The monopoly's objective is to maximize the revenue collected. The price $p_k$ is selected (using the optimal pricing policy) so as to influence or elicit the desired behavior (buy or not buy) from the social sensors. \\
A modification of (\ref{eq:Mon_Obj}) motivated by controlled information fusion applications in the presence of social learning is given by (\ref{eq:cst_es}) and (\ref{eq:RM}). 
Here $p_k$ is the incentive offered by the fusion center and $\Phi_s(k) \in (0,1)$ is the weight attached to the usefulness of the information acquired from sensor $k$. The objective of the information fusion is to maximize the number of sensors that act according to their observations, and estimate the underlying state. Since the sensors take into account the actions or decisions of the preceding sensors, fusion of informative decisions leads to improved estimate of the parameter, and hence improves the usefulness of information (in terms of reduction in the uncertainty of the Bayesian state estimate) fused by the fusion center and the successive sensors. 

 
\bibliographystyle{ieeetr}


\begin{IEEEbiography}[{\includegraphics[width=1in,height=1.25in,clip,keepaspectratio]{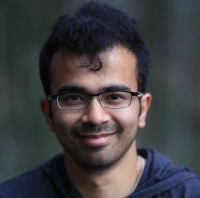}}]{Sujay Bhatt} received the Ph.D. degree in Electrical \& Computer Engineering from Cornell University, USA in 2019. He previously received the M. Tech degree in Electrical Engineering from Indian Institute of Technology, Bombay in 2014. His research interests are in network science, computational economics, statistical inference, stochastic control, and reinforcement learning.
\end{IEEEbiography}

\begin{IEEEbiography}[{\includegraphics[width=1in,height=1.25in,clip,keepaspectratio]{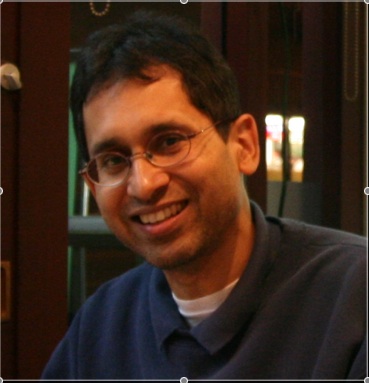}}]{Vikram Krishnamurthy} (F'05) received the Ph.D. degree from the Australian National University in 1992. He is currently a professor at Cornell Tech and the School of Electrical \& Computer Engineering, Cornell University. From 2002-2016 he was a Professor and Canada Research Chair at the University of British Columbia, Canada. His research interests include statistical signal processing and stochastic control in social networks and adaptive sensing. He served as Distinguished Lecturer for the IEEE Signal Processing Society and Editor-in-Chief of the IEEE Journal on Selected Topics in Signal Processing. In 2013, he was awarded an Honorary Doctorate from KTH (Royal Institute of Technology), Sweden. He is author of the books {\em Partially Observed Markov Decision Processes} and {\em Dynamics of Engineered Artificial Membranes and Biosensors} published by Cambridge University Press in 2016 and 2018, respectively.
\end{IEEEbiography}

\end{document}